\documentclass[aps,prb,twocolumn,showpacs,floatfix]{revtex4}

\usepackage{amsmath,graphicx}

\newcommand{\up}{\uparrow}
\newcommand{\dn}{\downarrow}

\begin{document}
\preprint{Submitted to Phys. Rev. B}


\title{Electronic States and Cyclotron Resonance in $n$-type InMnAs}

\author{G. D. Sanders}
\author{Y. Sun}
\author{F. V. Kyrychenko}
\author{C. J. Stanton}
\affiliation{Department of Physics, University of Florida, Box 118440\\
Gainesville, Florida 32611-8440}

\author{G. A. Khodaparast}
\author{M. A. Zudov}
\thanks{Present address: Physics Department, University of Utah,
Salt Lake City, Utah 84112, U.S.A.}
\author{J. Kono}
\affiliation{ Department of Electrical and Computer Engineering,
Rice University,\\ Houston, Texas 77005}

\author{Y. H. Matsuda}
\thanks{Present address: Department of Physics, Faculty of Science,
Okayama University, Okayama, Japan.}
\author{N. Miura}
\affiliation{Institute for Solid State Physics, University of Tokyo,
Kashiwanoha, Kashiwa, Chiba 277-8581, Japan}

\author{H. Munekata}
\affiliation{Imaging Science and Engineering Laboratory, Tokyo
Institute of Technology, Yokohama, Kanagawa 226-8503,
Japan}

\date{\today}


\begin{abstract}
We present a theory for electronic and magneto-optical properties
of $n$-type In$_{1-x}$Mn$_{x}$As magnetic alloy semiconductors in
a high magnetic field, $B \parallel \hat{z}$. We use an 8-band
Pidgeon-Brown model generalized to include the wavevector ($k_z$)
dependence of the electronic states as well as $s$-$d$ and $p$-$d$
exchange interactions with localized Mn $d$-electrons. Calculated
conduction-band Landau levels exhibit effective masses and $g$
factors that are strongly dependent on temperature, magnetic
field, Mn concentration ($x$), and $k_z$. Cyclotron resonance 
(CR) spectra are computed using
Fermi's golden rule and compared with ultrahigh-magnetic-field
($>$ 50 T) CR experiments, which show that the electron CR peak
position is sensitive to $x$. Detailed comparison between theory
and experiment allowed us to extract the $s$-$d$ and $p$-$d$
exchange parameters, $\alpha$ and $\beta$. We find that not only
$\alpha$ but also $\beta$ affects the electron mass because of the
strong interband coupling in this narrow gap semiconductor. In
addition, we derive analytical expressions for effective masses
and $g$ factors within the 8-band  model. Results indicates that
($\alpha - \beta$) is the crucial parameter that determines the
exchange interaction correction to the cyclotron masses. These
findings should be useful for designing novel devices based on
ferromagnetic semiconductors.
\end{abstract}



\maketitle

\section{Introduction}
\label{Introduction}

Recently, there has been much interest in III-V magnetic semiconductors
such as InMnAs\cite{Munekata89.1849} and GaMnAs.\cite{Ohno96.363}
The ferromagnetic exchange coupling between Mn ions
in these semiconductors is believed to be
mediated by free holes that are provided by Mn acceptors.
They become ferromagnetic at low temperatures and high enough
Mn concentrations.  Recent innovative
experiments have demonstrated the feasibility of controlling
ferromagnetism in these systems by tuning the carrier density
optically \cite{Koshihara97.4617} and
electrically.\cite{Ohno00.944}  Understanding their electronic,
transport, and optical properties is crucial for designing novel
ferromagnetic semiconductor devices with high Curie temperatures.
However, basic band parameters such as effective masses and $g$ factors
have not been accurately determined.
InMnAs alloys and their heterostructures with AlGaSb, the first grown
III-V magnetic semiconductor,
\cite{Munekata89.1849,Ohno92.2664,Munekata93.2929}
serve as a
prototype for implementing electron and hole spin degrees of freedom in
semiconductor spintronic devices.

The localized Mn spins strongly influence the delocalized conduction
and valence band states through the {\em s-d} and {\em p-d} exchange
interactions.  These interactions are usually parameterized as $\alpha$
and $\beta$, respectively.\cite{Furdyna88.29} Determining these
parameters is important for understanding the nature of Mn electron
states and their mixing with delocalized carrier states. In narrow gap
semiconductors like InMnAs, due to strong interband mixing, $\alpha$
and $\beta$ can influence {\it both} the conduction and valence bands.
This is in contrast to wide gap semiconductors where $\alpha$
influences primarily the conduction band and $\beta$ the valence band.
In addition, in narrow gap semiconductors, due to the strong interband
mixing, the coupling to the Mn spins does not affect all Landau levels
by the same amount.  As a result, the electron cyclotron resonance (CR)
peak can shift as a function of the Mn concentration, $x$. This can be
a sensitive method for estimating these exchange parameters.
For example, a recent CR study on Cd$_{1-x}$Mn$_x$Te
\cite{Matsuda02.115202} showed that the electron mass is strongly
affected by $sp$-$d$ hybridization.

In a recent Rapid Communication,\cite{Zudov02.161307} we reported the
first observation of electron CR in $n$-type In$_{1-x}$Mn$_x$As films
and described the dependence of cyclotron mass on $x$, ranging from
0 to 12\%. We observe that the electron CR peak shifts to lower field
with increasing $x$. Midinfrared interband absorption spectroscopy
revealed no significant $x$-dependence of the band gap.  A detailed
comparison of experimental results with calculations based on a
modified Pidgeon-Brown model allowed us to estimate $\alpha$ and
$\beta$ to be 0.5 eV and $-$1.0 eV, respectively.

In this paper, we describe details of the theoretical model and
comparison with the experiments. We use an 8-band Pidgeon-Brown
model, which is generalized to include the wavevector ($k_z$)
dependence of the electronic states as well as $s$-$d$ and $p$-$d$
exchange interactions with localized Mn $d$-electrons. Calculated
conduction-band Landau levels exhibit effective masses and $g$
factors that are strongly dependent on temperature, magnetic
field, Mn doping $x$, and $k_z$. At low temperatures and high $x$,
the sign
of the $g$ factor is positive and its magnitude exceeds 100. CR
spectra are computed using Fermi's golden rule. We also derive
analytical expressions for effective masses and $g$ factors within
the 8-band  model, which indicates that ($\alpha - \beta$) is the
crucial parameter that determines the exchange interaction
correction to the cyclotron masses. These findings should be
useful for desiging novel devices based on ferromagnetic
semiconductors.

\section{Experiment}
\label{Experiment}

We studied four $\sim$2-$\mu$m-thick In$_{1-x}$Mn$_x$As films with
Manganese concentrations
$x$ = 0, 0.025, 0.050 and 0.120 by ultrahigh-field magneto-absorption
spectroscopy.  The films were grown by low temperature molecular beam
epitaxy on semi-insulating GaAs substrates at 200$^{\circ}$C.  All the
samples were $n$-type and did not show ferromagnetism down to 1.5 K.
The electron densities and mobilities deduced from Hall measurements
are listed in Table \ref{table1}, together with the electron cyclotron
masses obtained at a photon energy of 117 meV (or a wavelength of
10.6 $\mu$m).

\begin{table}[bth]
\caption{Densities, mobilities, and cyclotron masses for the four
samples studied.  The densities and mobilities are in units of
cm$^{-3}$ and cm$^2$/Vs, respectively. The masses were obtained at a
photon energy of 117 meV (or $\lambda$ = 10.6 $\mu$m).}
\label{table1}
\begin{ruledtabular}
\begin{tabular*}{\hsize}{l@{\extracolsep{0ptplus1fil}}c@{\extracolsep{
0ptplus1fil}}c@{\extracolsep{0ptplus1fil}}c@{\extracolsep{0ptplus1fil}}
c} Mn content $x$ & 0 & 0.025 & 0.050 & 0.120 \\
\colrule
Density (4.2 K) & $\sim$1.0$\times$10$^{17}$ & 1.0$\times$10$^{16}$ &
0.9$\times$10$^{16}$ & 1.0$\times$10$^{16}$ \\
Density (290 K) & $\sim$1.0$\times$10$^{17}$ & 2.1$\times$10$^{17}$ &
1.8$\times$10$^{17}$ & 7.0$\times$10$^{16}$ \\
Mobility (4.2 K) & $\sim$4000 & 1300 & 1200 & 450 \\
Mobility (290 K) & $\sim$4000 & 400 & 375 & 450 \\
$m$/$m_0$ (30 K) & 0.0342 & 0.0303 & 0.0274 & 0.0263 \\
$m$/$m_0$ (290 K) & 0.0341 & 0.0334 & 0.0325 & 0.0272 \\
\end{tabular*}
\end{ruledtabular}
\end{table}

The single-turn coil technique \cite{Nakao85.1018} was used to generate
ultrahigh magnetic fields with a pulse duration of $\sim$7 $\mu$s.
The sample and pick-up coil were placed in a liquid helium flow
cryostat.  The single-turn coil breaks in the
outward direction, leaving the sample, the pick-up coil, and the
cryostat intact, which allows us to repeat such destructive pulsed
measurements on the same sample.  We used the 10.6 $\mu$m line
from a CO$_2$ laser and produced circular polarization using a CdS
quarter-wave plate.  The transmitted radiation was detected by a
fast HgCdTe photovoltaic detector.
Signals from the detector and pick-up coil were
transmitted via optical fiber to a multi-channel digitizer
located in a shielded measurement room.

\begin{figure}
\includegraphics[scale=1.00]{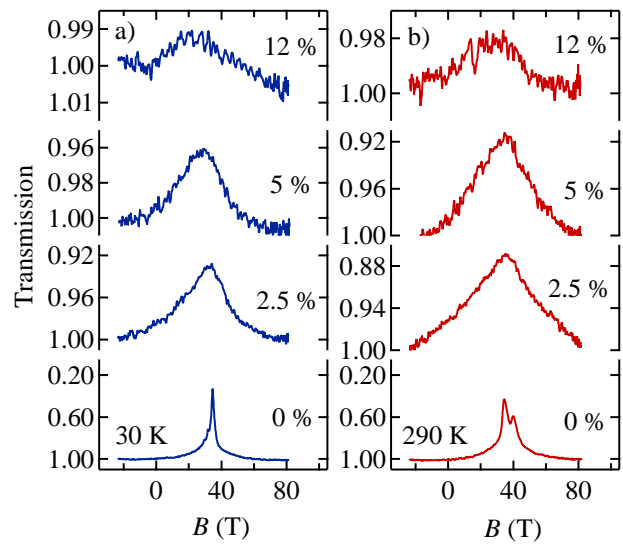}
\caption{Experimental CR spectra for different Mn concentrations, $x$,
taken at 290 (a) and 30 K (b).  The wavelength of the laser was
fixed at 10.6 $\mu$m with electron-active circular polarization
while the magnetic field $B$ was swept.
The resonance position shifts to lower $B$ with increasing $x$.
The $x$ values are 0\%, 2.5\%, 5\% and 12\%.}
\label{figcr}
\end{figure}

Typical measured CR spectra at 30 K and 290 K are shown in Figs.
\ref{figcr} (a) and (b), respectively.  Note that to compare the
transmission with absorption calculations, the transmission
increases in the negative $y$ direction.  Each figure shows
spectra for all four samples labelled by the corresponding Mn
compositions from 0 to 12 \%.  All the samples show pronounced
absorption peaks (or transmission dips) and the resonance field
{\em decreases} with increasing $x$.  Increasing $x$ from 0 to 12
\% results in a $\sim$25 \% decrease in cyclotron mass (see Table
\ref{table1}).  It is important to note that at resonance, the
densities and fields are such that only the lowest Landau level
for each spin type is occupied (see Fig. \ref{fig3}).  Thus, all
the electrons were in the lowest Landau level for a given spin
even at room temperature, precluding any density-dependent mass
due to nonparabolicity (expected at zero or low magnetic fields)
as the cause of the observed trend.

At high temperatures [e.g., Fig. \ref{figcr}(b)] the $x$ = 0 sample
clearly shows nonparabolicity-induced CR spin-splitting with the weaker
(stronger) peak originating from the lowest spin-down (spin-up) Landau
level, while the other three samples do not show such splitting.  The
reason for the absence of splitting in the Mn-doped samples is a
combination of 1) their low mobilities (which lead to substantial
broadening) and 2) the large effective $g$ factors due to the Mn
ions; especially in samples with large $x$ only the spin-down
level is substantially thermally populated (cf. Fig. \ref{fig3}).

\begin{figure}
\includegraphics[scale=1.00]{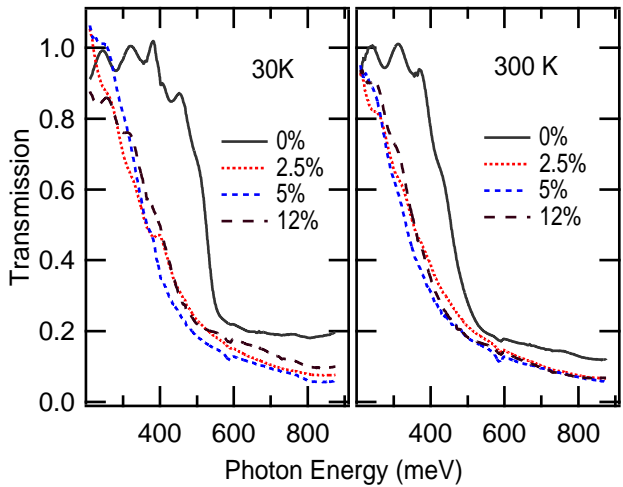}
\caption{Midinfrared transmission spectra
for the four samples studied, taken at 30 K and 300 K. The shift in band
gap for the $x=0$ sample can be attributed to a pronounced Burstein-Moss
shift resulting from the large electron density in the $x=0$ sample.}
\label{figftir}
\end{figure}

We also performed midinfrared interband absorption measurements at
various temperatures using Fourier-transform infrared spectroscopy to
determine how the band gap changes with Mn doping $x$.  In Fig.
\ref{figftir} we show transmission spectra for the four samples at two
temperatures (30 K and 300 K).  While a shift can be seen in the
bandgap in the $x=0$ samples,  we attribute this to a pronounced
Burstein-Moss shift resulting from the large electron density of the
$x$ = 0 sample. (As one increase Mn doping, the electron density
diminishes rapidly since the Mn ions act as acceptors).  The Fermi
energy for an electron density of 1 $\times$ 10$^{18}$ cm$^{-3}$ with a
mass of 0.023$m_0$ is 135 meV (without including the nonparabolicity).
This roughly accounts for the observed large shift in the absorption
edge.  Thus, our experimental data indicates that band gap does not
depend strongly on the Manganese concentration, $x$.

\section{Theory}
\label{Theory}

In this section we describe our effective mass theory of the electronic
and optical properties of $n$-type In$_{1-x}$Mn$_{x}$As alloys.
Our method is based on the Pidgeon-Brown \cite{Pidgeon66.575}
effective mass model of narrow gap semiconductors in a magnetic
field which includes the conduction electrons, heavy holes,
light holes, and split-off holes for a total of 8 bands when spin is
taken into account. In the original Pidgeon-Brown paper, only the zone
center $k_z=0$ states were considered and magnetic impurity effects
were not included.

The 8-band Pidgeon-Brown model can be generalized to
include the wavevector ($k_z$) dependence of the electronic states as
well as the $s$-$d$ and $p$-$d$ exchange interactions with localized
Mn $d$-electrons.\cite{Kossut88.183}

The magneto-optical  and cyclotron absorption can then be computed
using electronic states obtained from the generalized Pidgeon-Brown
model and Fermi's golden rule.  We will describe this in detail below.

\subsection{Effective mass Hamiltonian}

Following Pigeon and Brown \cite{Pidgeon66.575} we find it
convenient to separate the 8 Bloch basis states into an upper and
lower set which decouple at the zone center, i.e. $k_z=0$.
The Bloch basis states for the upper set are
\begin{subequations}
\label{upperset}
\begin{eqnarray}
\arrowvert 1 \rangle &=&
\arrowvert \frac{1}{2}, +\frac{1}{2} \ \rangle=
\arrowvert S \uparrow \rangle
\\
\arrowvert 2 \rangle &=&
\arrowvert \frac{3}{2}, +\frac{3}{2} \ \rangle=
\frac{1}{\sqrt{2}} \arrowvert (X + i Y) \uparrow \rangle
\\
\arrowvert 3 \rangle &=&
\arrowvert \frac{3}{2}, -\frac{1}{2} \rangle=
\frac{1}{\sqrt{6}} \arrowvert (X - i Y) \uparrow +2 Z \downarrow \rangle
\\
\arrowvert 4 \rangle &=&
\arrowvert \frac{1}{2}, -\frac{1}{2} \rangle=
\frac{i}{\sqrt{3}} \arrowvert -(X - i Y) \uparrow +Z \downarrow \rangle
\end{eqnarray}
\end{subequations}
which correspond to electron spin up, heavy hole spin up,
light hole spin down, and split off hole spin down.
Likewise, the Bloch basis states for the lower set are
\begin{subequations}
\label{lowerset}
\begin{eqnarray}
\arrowvert 5 \rangle &=&
\arrowvert \frac{1}{2},-\frac{1}{2} \rangle=
\arrowvert S \downarrow \rangle
\\
\arrowvert 6 \rangle &=&
\arrowvert \frac{3}{2}, -\frac{3}{2} \ \rangle=
\frac{i}{\sqrt{2}} \arrowvert (X - i Y) \downarrow \rangle
\\
\arrowvert 7 \rangle &=&
\arrowvert \frac{3}{2}, +\frac{1}{2} \rangle=
\frac{i}{\sqrt{6}} \arrowvert (X + i Y) \downarrow -2 Z \uparrow \rangle
\\
\arrowvert 8 \rangle &=&
\arrowvert \frac{1}{2}, +\frac{1}{2} \rangle=
\frac{1}{\sqrt{3}} \arrowvert (X + i Y) \downarrow +Z \uparrow \rangle
\end{eqnarray}
\end{subequations}
corresponding to electron spin down, heavy hole spin down,
light hole spin up, and split off hole spin up.

The effective mass Hamiltonian in bulk zinc blende materials
is given explicitly in Ref. \onlinecite{Efros88.7120}.
In the presence of a uniform magnetic field $B$ oriented along the
$z$ axis, the wave vector, $\vec{k}$, in the effective mass
Hamiltonian is replaced by the operator
\begin{equation}
\vec{k} = \frac{1}{\hbar} \left( \vec{p} + \frac{e}{c} \vec{A} \right)
\end{equation}
where $\vec{p}=-i \hbar \ \vec{\nabla}$ is the momentum operator
and $\vec{A}$ is the vector potential.
For the vector potential, we choose the Landau gauge
\begin{equation}
\vec{A} = - B \ y \ \hat{x}
\label{Gauge}
\end{equation}
from which we obtain
$\vec{B} = \vec{\nabla} \times \vec{A} = B \ \hat{z}$.

We introduce the appropriate creation and destruction operators
\begin{subequations}
\label{adaggera}
\begin{equation}
a^{\dagger} = \frac{\lambda}{\sqrt{2}} \left( k_x + i k_y  \right) ,
\label{creation}
\end{equation}
and
\begin{equation}
a = \frac{\lambda}{\sqrt{2}} \left( k_x - i k_y  \right) .
\label{destruction}
\end{equation}
\end{subequations}
The magnetic length, $\lambda$, is
\begin{equation}
\lambda = \sqrt{\frac{\hbar c}{e B}}
        = \sqrt{\frac{\hbar^2}{2 m_0} \ \frac{1}{\mu_B B}}.
\label{lambda}
\end{equation}
where $\mu_B=5.789 \times 10^{-5}\ \mbox{eV/Tesla}$ is the Bohr
magneton and $m_0$ is the free electron mass. Using Eqs.
(\ref{creation}) and (\ref{destruction}) to eliminate the
operators $k_x$ and $k_y$ in the effective mass Hamiltonian, we
arrive at the Landau Hamiltonian
\begin{equation}
H_L = \left[
\begin{array}{cc}
L_{a}           & L_{c} \\
L_{c}^{\dagger} & L_{b}
\end{array}
\right]
\label{H_L}
\end{equation}
with the submatrices $L_a$, $L_b$ and $L_c$ given by
\begin{equation}
L_a=\left[
\begin{array}{cccc}
E_g+A &
i \frac{V}{\lambda} a^{\dagger} &
i \sqrt{\frac{1}{3}} \frac{V}{\lambda} a &
  \sqrt{\frac{2}{3}} \frac{V}{\lambda} a \\
-i \frac{V}{\lambda} a &
-P-Q &
-M &
i\sqrt{2}M \\
-i \sqrt{\frac{1}{3}} \frac{V}{\lambda} a^{\dagger} &
-M^{\dagger} &
-P+Q &
i \sqrt{2} Q \\
\sqrt{\frac{2}{3}} \frac{V}{\lambda} a^{\dagger} &
-i \sqrt{2} M^{\dagger} &
-i \sqrt{2} Q &
-P-\Delta
\end{array}
\right]
\end{equation}
\begin{equation}
L_b=\left[
\begin{array}{cccc}
E_g+A &
- \frac{V}{\lambda} a &
-\sqrt{\frac{1}{3}} \frac{V}{\lambda} a^{\dagger} &
i \sqrt{\frac{2}{3}} \frac{V}{\lambda} a^{\dagger} \\
-\frac{V}{\lambda} a^{\dagger} &
-P-Q &
-M^{\dagger} &
i \sqrt{2} M^{\dagger} \\
-\sqrt{\frac{1}{3}} \frac{V}{\lambda} a &
-M &
-P+Q &
i \sqrt{2} Q \\
-i\sqrt{\frac{2}{3}} \frac{V}{\lambda} a &
-i \sqrt{2} M &
-i \sqrt{2} Q &
-P-\Delta
\end{array}
\right]
\end{equation}
\begin{equation}
L_c=\left[
\begin{array}{cccc}
0 &
0 &
\sqrt{\frac{2}{3}} V k_z   &
i \sqrt{\frac{1}{3}} V k_z \\
0 &
0 &
-L &
-i \sqrt{\frac{1}{2}} L \\
-i \sqrt{\frac{2}{3}} V k_z &
L &
0 &
i \sqrt{\frac{3}{2}} L^{\dagger} \\
-\sqrt{\frac{1}{3}} V k_z &
-i \sqrt{\frac{1}{2}} L &
i \sqrt{\frac{3}{2}} L^{\dagger} &
0
\end{array}
\right]
\end{equation}
In Eq. (\ref{H_L}), $E_g$ is the bulk band gap, and
$\Delta$ is the spin-orbit splitting. The Kane momentum matrix element
$V = -\frac{i}{\hbar} \langle S \arrowvert p_x \arrowvert X \rangle$
is related to the optical matrix parameter, $E_p$, by
\cite{Vurgaftman01.5815}
\begin{equation}
V = \sqrt{\frac{\hbar^2}{m_0} \ \frac{E_p}{2}}
\label{V}
\end{equation}
The operators $A$, $P$, $Q$, $L$ and $M$ are
\begin{subequations}
\label{APQLM}
\begin{equation}
A = \frac{\hbar^2}{m_0} \frac{\gamma_4}{2}
\left( \frac{2 N + 1}{\lambda^2} + k_z^2\right),
\end{equation}
\begin{equation}
P = \frac{\hbar^2}{m_0} \frac{\gamma_1}{2}
\left( \frac{2 N + 1}{\lambda^2} + k_z^2 \right),
\end{equation}
\begin{equation}
Q = \frac{\hbar^2}{m_0} \frac{\gamma_2}{2}
\left( \frac{2 N + 1}{\lambda^2} -2 k_z^2 \right),
\end{equation}
\begin{equation}
L = \frac{\hbar^2}{m_0} \gamma_3
\left( \frac{ -i \sqrt{6}\ k_z a}{\lambda} \right),
\end{equation}
and
\begin{equation}
M = \frac{\hbar^2}{m_0} \left( \frac{\gamma_2+\gamma_3}{2} \right)
\left(
\frac{\sqrt{3}}{\lambda^2} a^2
\right).
\end{equation}
\end{subequations}
In Eq. (\ref{APQLM}e), we have neglected a second term in $M$
proportional 
to $(\gamma_2-\gamma_3) (a^{\dagger})^2$.  We do this for two reasons:
1) $(\gamma_2-\gamma_3)$ is small and 2) this term will couple different
Landau manifolds making it more difficult to diagonalize the
Hamiltonian. The effect of this term can be accounted for later by 
perturbation theory.   

In Eq. (\ref{APQLM}), the number operator $N = a^{\dagger} a$.
The parameters $\gamma_1$, $\gamma_2$, and $\gamma_3$
used here are not the usual Luttinger parameters since this is 
an 8-band model, but instead are related to
the usual Luttinger parameters $\gamma_1^L$, $\gamma_2^L$,
and $\gamma_3^L$ through the relations \cite{Luttinger56.1030}
\begin{equation}
\gamma_1 = \gamma_1^L - \frac{E_p}{3 E_g},
\end{equation}
\begin{equation}
\gamma_2 = \gamma_2^L - \frac{E_p}{6 E_g},
\end{equation}
and
\begin{equation}
\gamma_3 = \gamma_3^L - \frac{E_p}{6 E_g}.
\end{equation}
This takes into account the additional coupling of the valence bands to
the conduction band not present in the six band  Luttinger model.

The parameter $\gamma_4$ is related to the conduction
band electron effective mass, $m_e^{*}$, through the relation
\cite{Efros88.7120}
\begin{equation}
\label{gamma4}
\gamma_4 =
\frac{1}{m_e^{*}} - \frac{E_p}{3} \left(
\frac{2}{E_g} + \frac{1}{E_g+\Delta}
\right)
\end{equation}

Note that $\gamma_4$ can be related to Kane paramater $F$
\begin{equation}
\label{KaneF}
F=\frac{1}{m_o}\sum_r \frac{\arrowvert\langle S\arrowvert p_x\arrowvert
u_r\rangle  \arrowvert^2}{(E_c-E_r)}
\end{equation}
by $\gamma_4 = 1+2F$. In Eq. (\ref{KaneF}), 
$\langle S\arrowvert p_x\arrowvert u_r \rangle$ is the momentum matrix
element between the $s$-like conduction bands with
energies near $E_c$ and remote bands $r$ with characteristic energies
$E_r$. The Kane parameter $F$ takes into account the higher, remote band
contributions to the  conduction band through second-order perturbation 
theory.\cite{Vurgaftman01.5815}

The Zeeman Hamiltonian is
\begin{equation}
H_Z=\frac{\hbar^2}{m_0}\frac{1}{\lambda^2} 
\left[ \begin{array}{cc} Z_a & 0 \\ 0   & -Z_a \end{array} 
\right] \label{H_Z} \end{equation}
where the $4 \times 4$ submatrix $Z_a$ is given by
\begin{equation}
Z_a=\left[
\begin{array}{cccc}
\frac{1}{2} & 0 & 0 & 0 \\
0 & -\frac{3}{2}\kappa & 0 & 0 \\
0 & 0 & \frac{1}{2}\kappa & -i\sqrt{\frac{1}{2}}(\kappa+1) \\
0 & 0 & i\sqrt{\frac{1}{2}}(\kappa+1) & \kappa + \frac{1}{2}
\end{array}
\right].
\label{Z_a}
\end{equation}
The value of $\kappa$ used in Eq. (\ref{Z_a}) is related to
$\kappa^L$ as defined by Luttinger through the
relation \cite{Luttinger56.1030}
\begin{equation}
\kappa = \kappa^L - \frac{E_p}{6 E_g}.
\end{equation}
For the Luttinger parameter, $\kappa^L$, we use the approximation
\cite{Dresselhaus55.368,Dresselhaus55.580,Pidgeon66.575}
\begin{equation}
\kappa^L=
\gamma_3^L+\frac{2}{3}\ \gamma_2^L-\frac{1}{3}\ \gamma_1^L-\frac{2}{3}.
\end{equation}

The exchange interaction between the Mn$^{++}$ $d$ electrons and
the conduction $s$ and valence $p$ electrons is treated in the
virtual crystal and molecular field approximation. The resulting
Mn exchange Hamiltonian is \cite{Kossut88.183}
\begin{equation}
H_{Mn}= x\ N_0\ \langle S_z \rangle \left[
\begin{array}{cc}
D_a & 0 \\
0   & -D_a
\end{array}
\right]
\label{H_Mn}
\end{equation}
where $x$ is the Mn concentration, $N_0$ is the number of
cation sites in the sample, and $\langle S_z \rangle$ is
the average spin on a Mn site. The $4 \times 4$ submatrix $D_a$
is
\begin{equation}
D_a= \left[
\begin{array}{cccc}
\frac{1}{2}\alpha & 0 & 0 & 0 \\
0 & \frac{1}{2}\beta & 0 & 0 \\
0 & 0 & -\frac{1}{6}\beta & -i\frac{\sqrt{2}}{3}\beta \\
0 & 0 & i\frac{\sqrt{2}}{3}\beta & \frac{1}{2}\beta
\end{array}
\right]
\label{D_a}
\end{equation}
where $\alpha$ and $\beta$ are the exchange integrals.

In the paramagnetic phase, the average spin on a Mn site is given
in the limit of noninteracting spins by
\begin{equation}
\langle S_z \rangle= -S\ B_S \left( g S \ \frac{\mu_B B}{kT} \right)
\label{S_z}
\end{equation}
where $g = 2$ and $S = \frac{5}{2}$ for
for the $3d^5$ electrons of the Mn$^{++}$ ion.\cite{Furdyna88.29}
The Brillouin function, $B_S(x)$, is defined as
\begin{equation}
B_S(x)=\frac{2S+1}{2S}\coth \left( \frac{2 S + 1}{2 S}x \right)
- \frac{1}{2S} \coth \left( \frac{x}{2S}\right)
\end{equation}
The antiparallel orientation of $B$ and $\langle S_z \rangle$
is due to the difference in sign of the magnetic moment and the
electron spin. Since $B$ is directed along the $z$ axis,
the average Mn spin saturates at $\langle S_z \rangle=-\frac{5}{2}$.

The total effective mass Hamiltonian for In$_{1-x}$Mn$_{x}$As in a
magnetic field directed along the $z$ axis is just the
sum of the Landau, Zeeman, and $sp$-$d$ exchange contributions, i.e.
\begin{equation}
H = H_L + H_Z + H_{Mn}.
\label{H_Total}
\end{equation}
We note that at $k_z=0$, the effective mass Hamiltonian is block
diagonal with respect to the upper and lower Bloch basis sets.

It is assumed in our calculations that the compensation arises from
As antisites and hence
the effective Mn fraction, $x$, in Eq. (\ref{H_Mn})
is taken to be equal to the actual Mn fraction in the sample.
We note that this is supported by experimantal evidence
showing that InAs grown at low temperature (200 $^{\circ}$C) is a
homogeneous alloy and that the magnetization varies linearly with
Mn content, $x$.
\cite{Munekata89.1849,Munekata90.176,Ohno91.6103,Molnar91.356}

\subsection{Energies and wavefunctions}

With the choice of gauge given in eq. (\ref{Gauge}), translational
symmetry in the $x$ direction is broken while translational symmetry
along the $y$ and $z$ directions is maintained. Thus, $k_y$ and $k_z$
are good quantum numbers and the envelope functions for the effective
mass Hamiltonian (\ref{H_Total}) can be written as
\begin{equation}
{\cal{F}}_{n,\nu} = \frac{e^{i(k_y y + k_z z)}}{\sqrt{{\cal{A}}}}
\left[
\begin{array}{l}
a_{n,1,\nu}(k_z) \ \phi_{n-1} \\
a_{n,2,\nu}(k_z) \ \phi_{n-2} \\
a_{n,3,\nu}(k_z) \ \phi_{n}   \\
a_{n,4,\nu}(k_z) \ \phi_{n}   \\
a_{n,5,\nu}(k_z) \ \phi_{n}   \\
a_{n,6,\nu}(k_z) \ \phi_{n+1} \\
a_{n,7,\nu}(k_z) \ \phi_{n-1} \\
a_{n,8,\nu}(k_z) \ \phi_{n-1}
\end{array}
\right]
\label{Fn}
\end{equation}

In Eq. (\ref{Fn}), $n$ is the Landau
quantum number associated with the Hamiltonian matrix,
$\nu$ labels the eigenvectors,
${\cal{A}} = L_x L_y$ is the cross sectional area of the sample in the
$xy$ plane, $\phi_n(\xi)$ are harmonic oscillator eigenfunctions
evaluated at $\xi = x - \lambda^2 k_y$, and $a_{n,m,\nu}(k_z)$ are
complex expansion coefficients for the $\nu$-th eigenstate which
depends explicitly on $n$ and $k_z$.  Note that the wavefunctions
themselves will be given by the envelope functions in Eq. {\ref{Fn}}
with each component multiplied by the corresponding
$k_z = 0$ Bloch basis states given
in Eqs. (\ref{upperset}) and (\ref{lowerset}).

Substituting ${\cal{F}}_{n,\nu}$ from Eq. (\ref{Fn})
into the effective mass Schr\"{o}dinger equation
with $H$ given by Eq. (\ref{H_Total}), we obtain a matrix eigenvalue
equation. By neglecting the second term in $M$ as described in
Eq. (\ref{APQLM}e), $H$ is block diagonal in the Landau quantum
number $n$. We obtain a set of matrix eigenvalue equations
\begin{equation}
H_n \ F_{n,\nu} = E_{n,\nu}(k_z) \ F_{n,\nu}.
\label{Schrodinger}
\end{equation}
which
can be solved separately for each allowed value of the Landau quantum
number, $n$, to obtain the Landau levels $E_{n,\nu}(k_z)$. The
components of the normalized eigenvectors, $F_{n,\nu}$, are the
expansion coefficients, $a_i$.

Since the harmonic oscillator functions, $\phi_{n'}(\xi)$, are only
defined for $n' \ge 0$, it follows from Eq. (\ref{Fn}) that
$F_{n,\nu}$ is defined for $n \ge -1$.
The energy levels are denoted $E_{n,\nu}(k_z)$ where $n$ labels the
Landau level and $\nu$ labels the eigenenergies belonging
to the same Landau level in ascending order.

For $n=-1$, we set all coefficients $a_i$ to zero except
for $a_6$ in order to prevent harmonic oscillator eigenfunctions
$\phi_{n'}(\xi)$ with $n' < 0$ from appearing in the wavefunction.
The eigenfunction in this case is a pure heavy hole spin down state and
the Hamiltonian is now a $1 \times 1$ matrix whose eigenvalue
corresponds to a heavy hole spin down Landau level.

For $n=0$, we must set $a_1=a_2=a_7=a_8=0$ and the Landau levels and
envelope functions are then obtained by diagonalizing a $4 \times 4$
Hamiltonian matrix obtained by striking out the appropriate rows and
columns. For $n=1$, the Hamiltonian matrix is $7 \times 7$ and
for $n \ge 2$ the Hamiltonian matrix is $8 \times 8$.

The matrix $H_n$ in Eq. (\ref{Schrodinger}) is the sum of Landau,
Zeeman, and exchange contributions. The explicit forms for the
Zeeman and exchange Hamiltonian matrices are given in Eqs. (\ref{H_Z})
and (\ref{H_Mn}) and are independent of $n$. The explicit form of the
Landau Hamiltonian for an arbitrary value of $n$ is given in
Appendix \ref{A}.

\subsection{Analytical solutions using Kane model}

In this section we derive analytical expressions that describe the
conduction band cyclotron resonance energies and $g$-factors in
In$_{1-x}$Mn$_x$As.

\subsubsection{General formalism}

The full Hamiltonian of the problem is given by Eq.~(\ref{H_Total}).
The natural way
to diagonalize this $8\times 8$ matrix analytically is to treat
the off-diagonal elements within perturbation theory. This
approach, however, remains valid only if the off-diagonal elements
are much smaller than the band gap (we are interested in the
conduction band states only). Due to strong $s$-$p$ coupling and
the relatively small band gap in InAs this condition breaks
down quickly with increasing magnetic field. In fact,
$V\lambda^{-1}$ exceeds the value of one half of band gap at $B
\approx$ 25 T.  

Since we are interested in ultrahigh field cyclotron resonance, 
we choose another approach similar to that used by
Kane.\cite{Kane57.249}  Namely, we neglect the small terms in the
Hamiltonian matrix (\ref{H_Total}), arising from the free electron
kinetic energy and the interaction with remote bands. In other
words, we neglect all terms proportional to $\lambda^{-2}$. Note
that these terms are small compared to the band gap. They are
proportional to the magnetic field and are on the order of $\sim
10 \ \mbox{meV}$ at a field of $B=100 \ \mbox{T}$.

Of course, one of the main drawbacks of the Kane
Hamiltonian is that the heavy-hole band is flat (neglecting the free
electron terms) and the Luttinger model as well as the modified
Pidgeon-Brown model discussed above, were developed to take into account
coupling to remote bands thereby giving the heavy hole band the correct
mass. Since in this work, we are only interested in the conduction
bands, this approximation is not crucial and it will allow us to obtain
analytical expressions for the conduction band energies.

Our Hamiltonian then is the sum of two parts. The first term is a
${\bf k\cdot p}$ Hamiltonian that takes into account only the
interactions between conduction and valence bands and the second term
is the carrier-magnetic ion exchange interaction. Neglecting the second
term, solutions of the Hamiltonian can be found analytically. For the
second term, even in the limit of saturated $\langle S_z \rangle$, the
exchange interaction is much smaller than the band gap and, thus, it
can be treated as a perturbation even in high magnetic fields.
Therefore, as unperturbed states we take the solutions of the
Kane-like Hamiltonian and consider the Mn~$s(p)$-$d$ exchange
interaction to first order in perturbation theory.

Solutions of the  Kane Hamiltonian can be written in the general form
\begin{equation}\label{Kane_transc}
  E=\frac1{m^*(E)}\left(\mu_B B(2n+1)+\frac{\hbar^2 k_z^2}{2
  m_0}\right)\pm \frac{1}2 g^*(E)\mu_B B,
\end{equation}
where we have introduced an energy dependent dimensionles effective mass
\begin{equation}\label{m(E)}
  \frac1{m^*(E)}=\frac{E_p}3\left[\frac2{E+E_g}+
  \frac1{E+E_g+\Delta}\right]
\end{equation}
and effective $g$ factor
\begin{equation}\label{g(E)}
  g^*(E)=\frac{2 E_p}3\left[\frac1{E+E_g+\Delta}-
  \frac1{E+E_g}\right].
\end{equation}
Note that we have neglected the free electron contributions to these
terms.
The upper and lower signs in Eq.~(\ref{Kane_transc})
correspond to spin up and spin down conduction band states,
respectively, and the zero of energy is chosen to lie at the bottom
of the conduction band when $B = 0$.

Note also, for this section only,  
we have redefined the index $n$ compared to the full
model which was discussed in the previous section.  In
Eq.~(\ref{Kane_transc}) for electron spin up we follow Pidgeon and
Brown \cite{Pidgeon66.575} and redefine the Landau quantum number $n$
by making the transformation $n \rightarrow n+1$ so that $n=0$
corresponds to the ground state Landau level for both spin up and spin
down solutions. While this convention on the numbering of $n$ is
convenient for discussing the Landau levels in the conduction band, it
can be applied only for states in the simplified  Kane Hamiltonian.
For the the more general  model where the index $n$ ranges from -1 to
$\infty$, the lowest spin up conduction Landau level is in the  $n=1$
manifold and the lowest lying spin down level is in the  $n=0$
manifold. Thus, our use of the Pidgeon and Brown labelling convention
for $n$ is confined to this section only.

Once solutions of Eq.~(\ref{Kane_transc}) are known, the first
order perturbative corrections due to the carrier-magnetic ion 
$s(p)$-$d$ exchange interaction is given by
\begin{widetext}
\begin{eqnarray}
  E^{(1)} & = & \pm\, \bar{x}N \left(\alpha + \beta \mu_B B
  \frac{E_p}9\left[
  \frac{8n+4\mp 5}{(E+E_g)^2}+\frac{2n+1\pm 1}{(E+E_g+\Delta)^2}+
  \frac{8n+4\pm 4}{(E+E_g)(E+E_g+\Delta)}\right] + \right. \nonumber \\
  \label{E1}
  & + & \left. \beta \frac{\hbar^2 k_z^2}{2 m_0} \frac{E_p}9\left[
  \frac2{(E+E_g)^2}-\frac1{(E+E_g+\Delta)^2}+
  \frac8{(E+E_g)(E+E_g+\Delta)}\right] \right),
\label{E1approx}
\end{eqnarray}
with
\begin{equation}\label{N}
 N=\left(1+\mu_B B \frac{E_p}3\left[\frac{4n+2\mp 1}{(E+E_g)^2}+
 \frac{2n+1\pm 1}{(E+E_g+\Delta)^2}\right]+ \frac{\hbar^2 k_z^2}{2 m_0}
 \frac{E_p}9\left[\frac2{(E+E_g)^2}+\frac1{(E+E_g+\Delta)^2}
 \right]   \right)^{-1},
\end{equation}
\end{widetext}
where we introduce the notation
$\bar{x}= x N_0 \langle S_z \rangle /2$.
Upper and lower signs again correspond to spin up and
spin down conduction band states respectively.

Note that from Eq. (\ref{E1}) we can immediately gain some insight
into the effects of the $s(p)$-$d$ exchange interaction on the narrow
gap material.
i) Since we do not take into account variation of the material
parameters
(such as the energy gap) with manganese concentration, $x$, it follows
immediately from Eq.~(\ref{E1}) that we obtain a linear dependence of
the conduction band energies on $x$. 
ii) Without conduction-valence band mixing, the first order
correction would be $E^{(1)}  =  \pm\, \bar{x}N \alpha$ as one might
expect.  (In this case, there would be no shift in the cyclotron
resonance energy with Mn doping $x$ since all levels for a given spin
would be shifted by the same amount).  The term proportional to $\beta$
is a direct consequence of conduction-valence band mixing.
iii) It is also seen from Eq.~(\ref{E1approx}) that both band mixing
contributions induced by the magnetic field, $B$, and motion in the
$z$-direction ($k_z$) have the same sign and, since $\alpha$ and
$\beta$ are of opposite signs, both reduce $E^{(1)}$.

To calculate the conduction band energy spectrum using
Eqs.~(\ref{E1})-(\ref{N}) we need to know the energies of the
unperturbed problem. In spite of its clear form,
Eq.~(\ref{Kane_transc}) is indeed a third order equation and its
general solutions are quite complicated. Standard approximations
usually assume either strong spin-orbit interaction ($\Delta\gg E_g+E$)
\cite{Mavroides72.351} or  small kinetic energy ($E\ll E_g+\Delta$).
\cite{Zawadzki91.483} In
the present paper we are interested in strong magnetic fields
where $E\sim E_g\approx \Delta$ and have to use the general
solution. However, in our situation we can make one more
simplification, suitable for the particular case of $\rm
InAs$-based semiconductors. Using the fact that in $\rm InAs$ the
energy gap is approximately equal to the spin-orbit splitting, we
set $E_g=\Delta$. Although this approximation is not necessary to
obtain analytical solutions, it makes the final expressions more
readable.

With these simplifications solutions of Eq.~(\ref{Kane_transc}),
i.e. energies of unperturbed states in our model, can be
presented as
\begin{equation}\label{Kane_E}
  E^{(0)}\approx E_0 + T_z,
\end{equation}
with
\begin{equation}\label{Tz}
  T_z=\frac1{m^*(E_0)}\frac{\hbar^2 k_z^2}{2 m_0}.
\end{equation}
The position of the bottom of the Landau subbands have the form
\begin{widetext}
\begin{equation}\label{E0}
  E_0 =
  2\sqrt{\frac{\Delta^2+(2n+1)V^{\prime \ 2}}{3}}\cos\left[\frac13
  \arccos\left(\frac{\sqrt{3}(4n+2\mp 1)\Delta
  V^{\prime \ 2}}{2\sqrt{[\Delta^2+(2n+1)V^{\prime \ 2}]^3}}\right)
  \right]-\Delta,
\end{equation}
\end{widetext}
where upper (lower) sign corresponds to electron spin up (down)
states. In Eq.~(\ref{E0}) we have introduced $V^{\prime}\equiv
V\lambda^{-1}=\sqrt{\mu_B B E_p}$.

The second term in Eq.~(\ref{Kane_E}) is the kinetic energy for
the motion along the magnetic field. It assumes parabolic
dispersion with the effective masses being different for each
Landau subband. This approximation is valid near the bottom of the
Landau subbands and is justified when $T_z \ll E_g+E_0$.

Equations~(\ref{E1})-(\ref{E0}) fully describe, within the Kane
model, the conduction band energy spectra of In$_{1-x}$Mn$_x$As in
a magnetic field. From the energy spectra one can easily calculate
different physical quantities observed in cyclotron resonance
experiments, such as cyclotron masses or $g$ factors. However, the
general expressions are rather complicated for direct analysis,
but can be significantly simplified in the limits of low
($V^{\prime}\ll\Delta$) and high ($V^{\prime}\gg\Delta$) magnetic
fields. First we consider the situation at the bottom of the
Landau subbands. Then we take into account finite $k_z$.

\subsubsection{Low-field and high-field limits}

The cyclotron energies are defined as the splitting between the
two lowest Landau levels ($n=0$ and $n=1$) and are given by
$\Delta E \equiv E(n=1)-E(n=0)$.

The low magnetic field limit of the cyclotron energies can be
obtained by expanding to second order in powers of
$V^{\prime}/\Delta \ll 1$. The result for spin up states is
\begin{equation}\label{CEal}
  \Delta E_{\up}\approx \Delta \frac53
  \frac{V^{\prime \ 2}}{\Delta^2}-\bar{x} (\alpha-\beta)
  \frac32\frac{V^{\prime \ 2}}{\Delta^2}-\bar{x}\beta\frac19
  \frac{V^{\prime \ 2}}{\Delta^2},
\end{equation}
and for spin down states we obtain
\begin{equation}\label{CEbl}
  \Delta E_{\dn}\approx \Delta \frac53
  \frac{V^{\prime \ 2}}{\Delta^2}+\bar{x} (\alpha-\beta)
  \frac32\frac{V^{\prime \ 2}}{\Delta^2}+\bar{x}\beta\frac19
  \frac{V^{\prime \ 2}}{\Delta^2}.
\end{equation}
It is seen that in small magnetic fields the cyclotron energy
splitting increases linearly with the field that corresponds to 
simple parabolic dispersions at small $k$.

Likewise, the cyclotron energy in the high magnetic field limit is
obtained by expanding the energy splitting to second order
in $\Delta/V^{\prime} \ll 1$. The results are
\begin{equation}\label{CEah}
   \Delta E_{\up}\approx V^{\prime}\left( \sqrt3-1
  +\frac19\frac{\Delta}{V^{\prime}}-
  0.24\frac{\Delta^2}{V^{\prime \ 2}}\right)-\bar{x}(\alpha-\beta)
  0.21\frac{\Delta}{V^{\prime}},
\end{equation}
for spin up and
\begin{equation}\label{CEbh}
  \Delta E_{\dn}\approx
  V^{\prime}\left(\sqrt3-1-\frac19\frac{\Delta}{V^{\prime}}+
  0.03\frac{\Delta^2}{V^{\prime \ 2}}\right)+\bar{x}(\alpha-\beta)0.07
  \frac{\Delta}{V^{\prime}},
\end{equation}
for spin down. At high magnetic fields, nonparabolicity due to
conduction and valence band mixing results in a square
root dependence of the cyclotron energy on magnetic field.
\cite{McCombe75.1,McCombe72.321,Mavroides72.351,Kinch71.461}

It is also seen from Eqs. (\ref{CEal})-(\ref{CEbh}) that the
electron energy mainly depends not on two independent exchange
constants $\alpha$ and $\beta$, but on their difference
($\alpha-\beta$). This feature will be discussed in more detail in
section \ref{cyclotron-mass}.

It is also easy to obtain limiting expressions for the spin
splitting of the conduction band states. For the lowest Landau
sublevels this spin splitting $\Delta E_0\equiv
E_{\up}(n=0)-E_{\dn}(n=0)$ is
\begin{equation}\label{dE0l}
  \Delta E_0\approx -\frac{\Delta}3\frac{V^{\prime \ 2}}{\Delta^2} +
  2\bar{x}\alpha-\bar{x}(\alpha-\beta)
  \frac32\frac{V^{\prime \ 2}}{\Delta^2}-
  \bar{x}\beta\frac19\frac{V^{\prime \ 2}}{\Delta^2},
\end{equation}
in the low magnetic field limit and
\begin{equation}\label{dE0h}
  \Delta E_0\approx -\frac{\Delta}3
  \left(1-\frac{\Delta}{V^{\prime}}\right)+
  \bar{x}(\alpha+\beta)+\bar{x}\frac23\frac{\Delta}{V^{\prime}}
  (\alpha-\beta)
\end{equation}
in the high-field limit. The first term in Eq.~(\ref{dE0l})
corresponds to a well-known low-field negative contribution to the
electron $g$ factor due to the influence of the spin-orbit split
valence bands $\Gamma_8$ and $\Gamma_7$ \cite{Roth59.90} and can
be obtained by putting $E_g=\Delta$ in the conventional expression
of Eq.~(\ref{g(E)}). The second term in Eq.~(\ref{dE0l})
is opposite in sign and describes a
contribution due to the $s$-$d$ exchange interaction. This term
dominates in low magnetic fields and is responsible for the giant
spin splitting of the conduction bands observed in diluted
magnetic semiconductors. With increasing magnetic field
$\langle S_z \rangle$ quickly saturates and the first term in
Eq.~(\ref{dE0l}) becomes dominant. Under certain conditions this
may result in a change of the sign of the spin splitting. If we go
beyond the low-field limit, the first term in Eq.~(\ref{dE0l})
ceases to be linear in the magnetic field and, finally, saturates
to $-\Delta/3$ in the high-field limit, as seen in
Eq.~(\ref{dE0h}). At the same time, since exchange constants
$\alpha$ and $\beta$ have the opposite signs, strong
valence-conduction band mixing results in a reduction of the exchange
induced spin splitting of the conduction band and a term
$\bar{x}(\alpha+\beta)$ appears in Eq.~(\ref{dE0h}) instead of
$2\bar{x}\alpha$ in Eq.~(\ref{dE0l}).

Finally, let us note that the spin splitting is different for
different Landau sublevels. For example, for the first excited
Landau sublevels (with $n=1$) the expressions analogous to
Eqs.~(\ref{dE0l}) and (\ref{dE0h}) have the form
\begin{equation}\label{dE1l}
  \Delta E_1\approx -\frac{\Delta}3\frac{V^{\prime \ 2}}{\Delta^2} +
  2\bar{x}\alpha-\bar{x}(\alpha-\beta)
  \frac92\frac{V^{\prime \ 2}}{\Delta^2}-
  \bar{x}\beta\frac13\frac{V^{\prime \ 2}}{\Delta^2}
\end{equation}
in the low-field limit and
\begin{equation}\label{dE1h}
  \Delta E_1\approx -\frac{\Delta}9\left(1-\frac1{\sqrt3}
  \frac{\Delta}{V^{\prime}}\right)+
  \bar{x}(\alpha+\beta)+\bar{x}\frac2{3\sqrt3}
  \frac{\Delta}{V^{\prime}}(\alpha-\beta)
\end{equation}
in the high-field limit.
There are two reasons for this behavior. First, with increasing
$n$ the energy separation between conduction and valence band
states increases and this suppresses the influence of the valence band
on the electron $g$ factor. As a result, in the high-filed limit, the
corresponding term saturates to $-\Delta/9$, which is three times
smaller than for Landau sublevels with $n=0$. Second, the Landau
quantum number $n$ affects the degree of conduction-valence
band mixing, resulting in a dependence of the exchange interaction
contribution to the $g$ factor on $n$.

\subsubsection{Finite $k_z$}

The main physics associated with finite $k_z$ is the presence of
additional conduction-valence band mixing induced by
free motion along the magnetic field. This effect is most
pronounced if we consider low-field spin splitting of the
conduction band states. Expanding the corresponding expressions to
first order in powers of $T_0/\Delta\ll 1$ ($T_0$ is the kinetic
energy (\ref{Tz}) with conduction band edge effective mass) we
find that the term
\begin{equation}\label{dE0lz}
  \Delta E_0^{k_z}\approx -\bar{x}\frac95(\alpha-\beta)
  \frac{T_0}{\Delta}-\bar{x}\frac4{15}\beta\frac{T_0}{\Delta}
\end{equation}
should be added to the Eq.~(\ref{dE0l}) in general case of states
with $k_z\neq 0$. The last two terms in Eq.~(\ref{dE0l}) together with
Eq.~(\ref{dE0lz}) describe the influence of the $s$-$p$ coupling on
the spin splitting of the conduction band Landau subbands. It is worth
noting that both these contributions lead to a reduction of the
$s$-$d$ exchange splitting $2\bar{x}\alpha$. Let us also note the
difference in their manifestation. Magnetic field induced $s$-$p$
mixing becomes pronounced at relatively strong magnetic fields. At
such fields, however, the role of exchange interaction is
decreased by the conventional spin splitting (first term in
Eq.~(\ref{dE0l})). At the same time, the $k_z$ induced contribution
could be important even in small magnetic fields, when the
exchange interaction plays the dominant role in spin splitting of
conduction band states.

In our model, exchange interaction is considered within first
order perturbation theory. As a result, all physical quantities
like $g$-factors or cyclotron masses are linear functions of the
manganese concentration $x$. This means, in particular, that if the
exchange interaction contribution to the spin splitting vanishes for a
certain concentration of magnetic ions, it should vanish at the
same $k_z$, magnetic field and temperature for any value of $x$.
The results of numerical diagonalization of the full Hamiltonian
matrix (\ref{H_Total}), presented in Fig.~\ref{gkzeps}, supports
this conclusion and justifies our model.

\subsection{Magneto-optical absorption}

In this section, we discuss how we calculate the magneto-optical
properties.
We calculate the  magneto-optical absorption coefficient at the photon
energy $\hbar \omega$ from
\cite{Bassani}
\begin{equation}
\alpha(\hbar \omega)=
\frac{\hbar \omega}{(\hbar c) n_r}\ \epsilon_2(\hbar \omega)
\end{equation}
where $\epsilon_2(\hbar \omega)$ is the imaginary part of the
dielectric function and $n_r$ is the index of refraction. The
imaginary part of the dielectric function is found using Fermi's
golden rule. The result is
\begin{eqnarray}
&&
\epsilon_2(\hbar \omega) =  \frac{e^2}{\lambda^2(\hbar \omega)^2}
\sum_{n,\nu; n',\nu'} \int_{-\infty}^{\infty} dk_z \
\arrowvert \hat{e} \cdot \vec{P}_{n,\nu}^{n',\nu'}(k_z) \arrowvert^2
\nonumber \\
&&
\times \left( f_{n,\nu}(k_z) - f_{n',\nu'}(k_z) \right)
\delta
\left( \Delta E_{n',\nu'}^{n,\nu}(k_z) - \hbar \omega \right),
\label{epsilon2}
\end{eqnarray}
where
$\Delta E_{n',\nu'}^{n,\nu}(k_z)=E_{n',\nu'}(k_z)-E_{n,\nu}(k_z)$
is the transition energy.
The function $f_{n,\nu}(k_z)$ in Eq. (\ref{epsilon2})
is the probability that the state
$(n,\nu,k_z)$, with energy $E_{n,\nu}(k_z)$, is occupied. It is
given by the Fermi distribution function
\begin{equation}
f_{n,\nu}(k_z) = \frac{1}{1+\exp[(E_{n,\nu}(k_z) - E_f)/kT]}.
\label{fnk}
\end{equation}

The Fermi energy $E_f$ in Eq. (\ref{fnk}) depends on temperature
and doping.
If $N_D$ is the donor concentration and $N_A$
the acceptor concentration, then the net donor
concentration, $N_C = N_D - N_A$, can be either positive
or negative depending on whether the sample is $n$- or $p$-type.
For a fixed temperature and Fermi level, the
net donor concentration is
\begin{equation}
N_C = \frac{1}{(2 \pi)^2 \lambda^2} \sum_{n,\nu}
\int_{-\infty}^{\infty} dk_z \ (f_{n,\nu}(k_z) - \delta_{n,\nu}^v)
\label{N_C}
\end{equation}
where $\delta_{n,\nu}^v = 1$ if the subband $(n,\nu)$ is
a valence band and vanishes if $(n,\nu)$ is a conduction band.
Given the net donor concentration and the temperature,
the Fermi energy can be found from Eq. (\ref{N_C}) using
a root finding routine.

Since the envelope functions and vector potential are slowly
varying over a unit cell, the dominant contributions to the optical
matrix elements are given by
\begin{eqnarray}
&&
\hat{e} \cdot \vec{P}_{n,\nu}^{n',\nu'}(k_z) =
\sum_{m,m'} a^{*}_{n,m,\mu}(k_z) \ a_{n',m',\mu'}(k_z)
\nonumber \\
&&
\times
\langle \phi_{N(n,m)} \arrowvert \phi_{N(n',m')} \rangle
\ \langle m \arrowvert ( \hat{e} \cdot \vec{P} ) \arrowvert m' \rangle
\label{EdotP}
\end{eqnarray}
where $\hat{e}$ is the unit polarization vector of the radiation,
$a_{n,m,\mu}(k_z)$ are the complex expansion coefficients for
the envelope functions in Eq. (\ref{Fn}), and
$\phi_{N(n,m)}$ are orthonormalized harmonic oscillator
wavefunctions. Their quantum numbers $N(n,m)$ depend explicitly on
$n$ and $m$ as defined in Eq. (\ref{Fn}).
In Eq. (\ref{EdotP}), we have neglected a term that depends on the
momentum matrix element, $\langle \phi_{N(n,m)} \arrowvert ( \hat{e}
\cdot \vec{P} ) \arrowvert\phi_{N(n',m')} \rangle $ between the
oscillator states.  Owing to strong band mixing in the narrow gap
materials, this term is much smaller than the momentum matrix elements
between the Bloch basis functions,
{\it even for intraband optical absorption} such
as for cyclotron resonance, hence we neglect it.

The momentum matrix elements
$\langle m \arrowvert P_x \arrowvert m' \rangle$,
$\langle m \arrowvert P_y \arrowvert m' \rangle$, and
$\langle m \arrowvert P_z \arrowvert m' \rangle$
are the momentum matrix elements between the Bloch basis functions
$\arrowvert m \rangle$ defined in
Eqs. (\ref{upperset}) and (\ref{lowerset}). Explicit expressions
for the momentum matrices $P_x$, $P_y$, and $P_z$ are found in
Appendix \ref{B}.

In our simulations, we consider e-active circularly polarized
light incident along the $z$ axis. For e-active circular
polarization, the unit polarization vector is
$\hat{e}=(\hat{x}-i\hat{y})/\sqrt{2}$. In performing the integral
in (\ref{epsilon2}) the Dirac delta function, $\delta(x)$, in
Fermi's golden rule is replaced by the Lorentzian lineshape
function $\Delta_{\gamma}(x)$ with full width at half maximum
(FWHM) of $\gamma$.

\subsection{Material parameters}

The material parameters we use are shown in Table
\ref{Parameters}. For the temperature dependent energy gap, $E_g$,
of InAs we use the empirical Varshni formula
\cite{Vurgaftman01.5815,Varshni67.149}
\begin{equation}
E_g=E_0-\frac{a\ T^2}{T+b}
\label{Egap}
\end{equation}
with $E_0=0.417\ \mbox{eV}$, $a=2.76 \times 10^{-4}\ \mbox{eV/K}$,
and $b=93\ \mbox{K}$.
As seen in Table I, the experimental cyclotron mass does not vary
significantly with temperature. Other low field
CR studies of InAs \cite{Palik61.131} also show a very weak
temperature dependence of the cyclotron mass. 
To account for a mass that only slightly varies with 
temperature, we keep the mass constant with 
temperature and adjust the $\gamma_4$ parameter (Eq. (\ref{gamma4})) 
to account for the temperature dependent band gap.
Alternatively, one might wish to keep
$\gamma_4$ constant and vary the ${\bf k}\cdot{\bf p}$ optical matrix
element $E_p$.

\begin{table}[tbp]
\caption{\label{Parameters}
Material parameters for In$_{1-x}$Mn$_{x}$As}
\begin{ruledtabular}
\begin{tabular}{lr}

Energy gap (eV) \footnotemark[1]                        & \\
$E_g$ ($T=30$  K)     &  0.415                            \\
$E_g$ ($T=290$ K)     &  0.356                            \\

Electron effective mass ($m_0$)                         & \\
$m_e^{*}$             &   0.022                           \\

Luttinger parameters \footnotemark[2]                   & \\
$\gamma_1^L$          &  20.0                             \\
$\gamma_2^L$          &   8.5                             \\
$\gamma_3^L$          &   9.2                             \\
$\kappa^L$            &  7.53                             \\

Spin-orbit splitting (eV) \footnotemark[2]              & \\
$\Delta$              &   0.39                            \\

Mn $s$-$d$ and $p$-$d$
exchange energies (eV)                                  & \\
$N_0\ \alpha$          & -0.5                             \\
$N_0\ \beta$           &  1.0                             \\

Optical matrix parameter (eV) \footnotemark[2]          & \\
$E_p$                 &  21.5                             \\

Refractive index \footnotemark[3]                       & \\
$n_r$                 &  3.42                             \\

\end{tabular}
\end{ruledtabular}

\footnotetext[1] {Eq. \ref{Egap}
with parameters from Ref.\ \onlinecite{Vurgaftman01.5815}. }
\footnotetext[2] {Ref.\ \onlinecite{Vurgaftman01.5815}. }
\footnotetext[3] {Ref.\ \onlinecite{Pankove}. }

\end{table}
%

\section{Results}
\label{n-Results}

\subsection{Landau levels and $g$ factors}

\begin{figure}[tbp]
\includegraphics[scale=0.48]{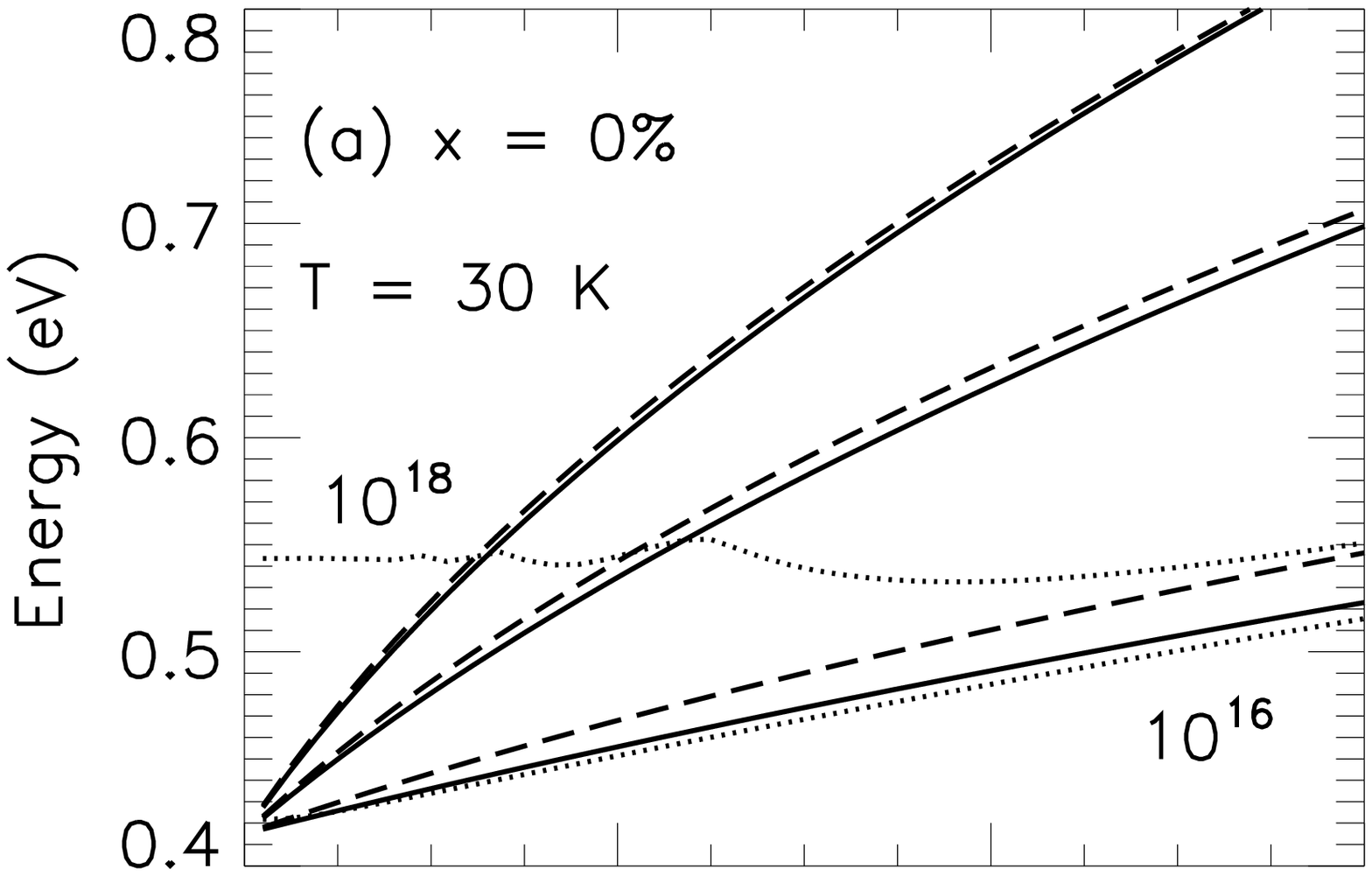}
\includegraphics[scale=0.48]{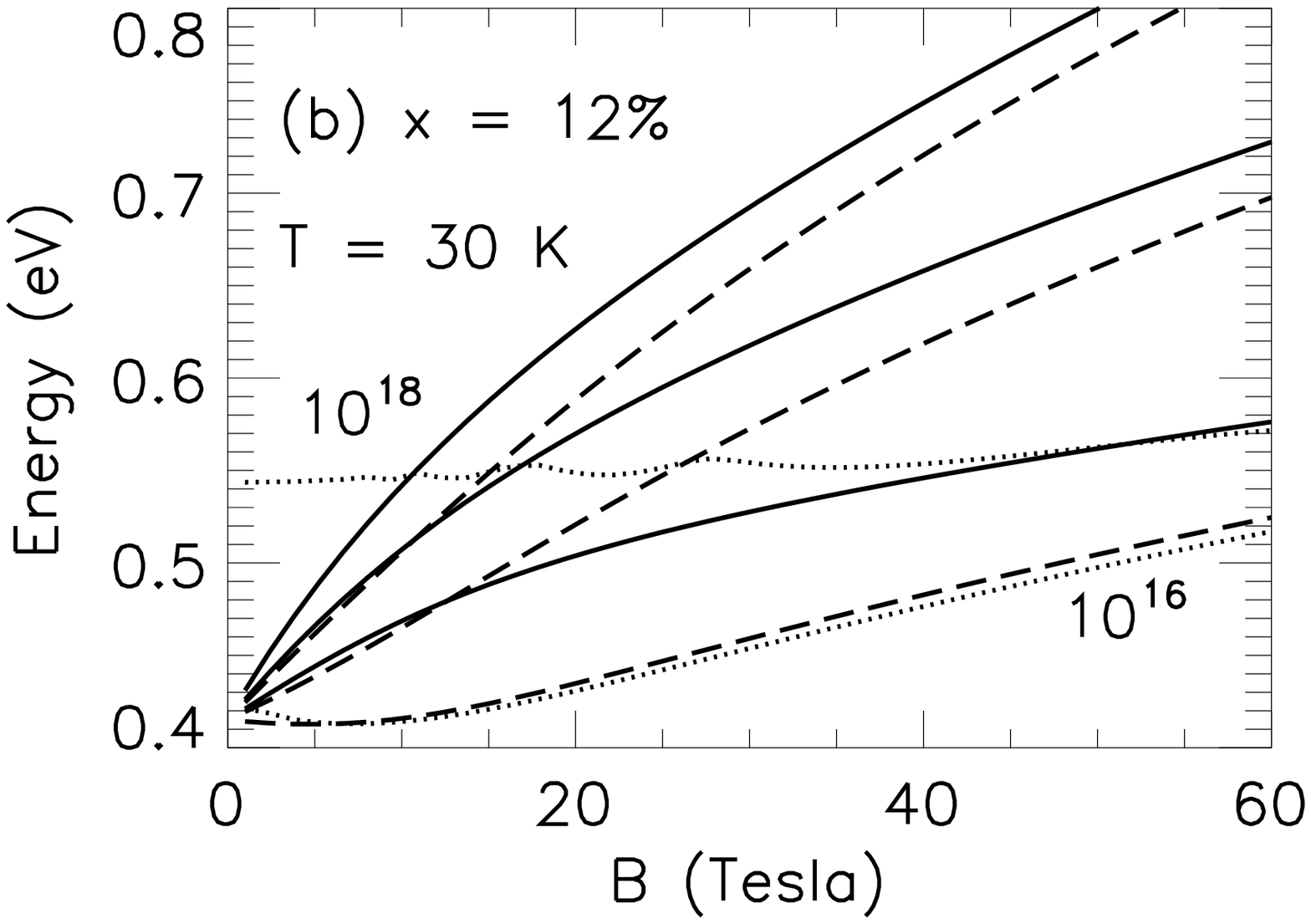}
\caption{\label{fig3}
Zone center Landau conduction subband energies at $T = 30 \ \mbox{K}$
as functions of magnetic field in $n$-doped In$_{1-x}$Mn$_{x}$As for
(a) $x=0\%$ and (b) $x=12\%$. Solid lines are spin-up and
dashed lines are spin-down levels. The Fermi energies are shown
as dotted lines for $n = 10^{16} \mbox{cm}^{-3}$ and
$n = 10^{18} \mbox{cm}^{-3}$. }
\end{figure}
%

In Fig. \ref{fig3} (a), the six lowest lying zone center ($k_z=0$)
Landau conduction subband energies in InAs are shown as functions
of applied magnetic field. At $k_z=0$, the submatrix, $L_c$, in
the Landau Hamiltonian, $H_L$, vanishes identically
and the Hamiltonian, $H$, in Eq. (\ref{H_Total}) is
block diagonal with respect to the upper and lower Bloch basis
states in Eqs. (\ref{upperset}) and (\ref{lowerset}).
In Fig. \ref{fig3} the spin-up Landau levels are shown as solid
lines while spin-down Landau levels are shown as dashed lines.
The Fermi energies for two different
electron concentrations, $n=10^{16}\ \mbox{cm}^{-3}$ and
$n=10^{18}\ \mbox{cm}^{-3}$, are shown as dotted lines. At
a concentration of $n=10^{16}\ \mbox{cm}^{-3}$, only the lowest
Landau subband is occupied at $B = 60$ T, while at
$n=10^{18}\ \mbox{cm}^{-3}$, the first two Landau subbands
are occupied.

In the absence of Mn impurities,
it is well known that at low magnetic fields, the spin splittings
between the electron spin up and spin down Landau levels, at the band
edge, can be described in terms of an effective gyromagnetic factor
\cite{Roth59.90,Rigaux88.229}
\begin{equation}
g^{*}=2 \left(1-\frac{E_p}{3 E_g}\frac{\Delta}{E_g+\Delta}\right)
\label{gstar}
\end{equation}
This is just Eq. (\ref{g(E)}) with $E=0$ and the {\it bare} {\it g}
factor 2 included.   
The $g$ factor in Eq. (\ref{gstar}) depends on temperature through the
temperature-dependent band gap. For bulk InAs at $T=30\ \mbox{K}$
we find $g^{*}=-14.7$ and for $T=290\ \mbox{K}$ we have
$g^{*}=-19.0$.
At nonzero magnetic fields, the expression for the gyromagnetic factor
in Eq. (\ref{gstar}) is not correct. As a function of the magnetic
field, the spin splittings become nonlinear as seen in Fig \ref{fig3}(a)
and depend explicitly on the Landau subband level indices, $n$.
This results from the nonparabolicity in the narrow gap material.

The effect of doping InAs with Mn is shown in Fig. \ref{fig3}(b)
where the Landau subband levels for In$_{0.88}$Mn$_{0.12}$As are
plotted.
At low fields, the effect of doping with Mn is to alter the
gyromagnetic factor. The gyromagnetic factor in
the presence of Mn impurities is \cite{Rigaux88.229}
\begin{equation}
g_{Mn}^{*}= g^{*}+\frac{x\ N_0 \alpha \ \langle S_z \rangle}{\mu_B B}
\label{gmnstar}
\end{equation}
Since the sign of $\alpha$ is negative, the gyromagnetic factor
increases with the applied magnetic field. At sufficiently high Mn
concentrations and magnetic fields, the low temperature
gyromagnetic factor can become positive and the spin splittings
reverse as seen in Fig. \ref{fig3}(b).

\begin{figure}[tbp]
\includegraphics[scale=0.48]{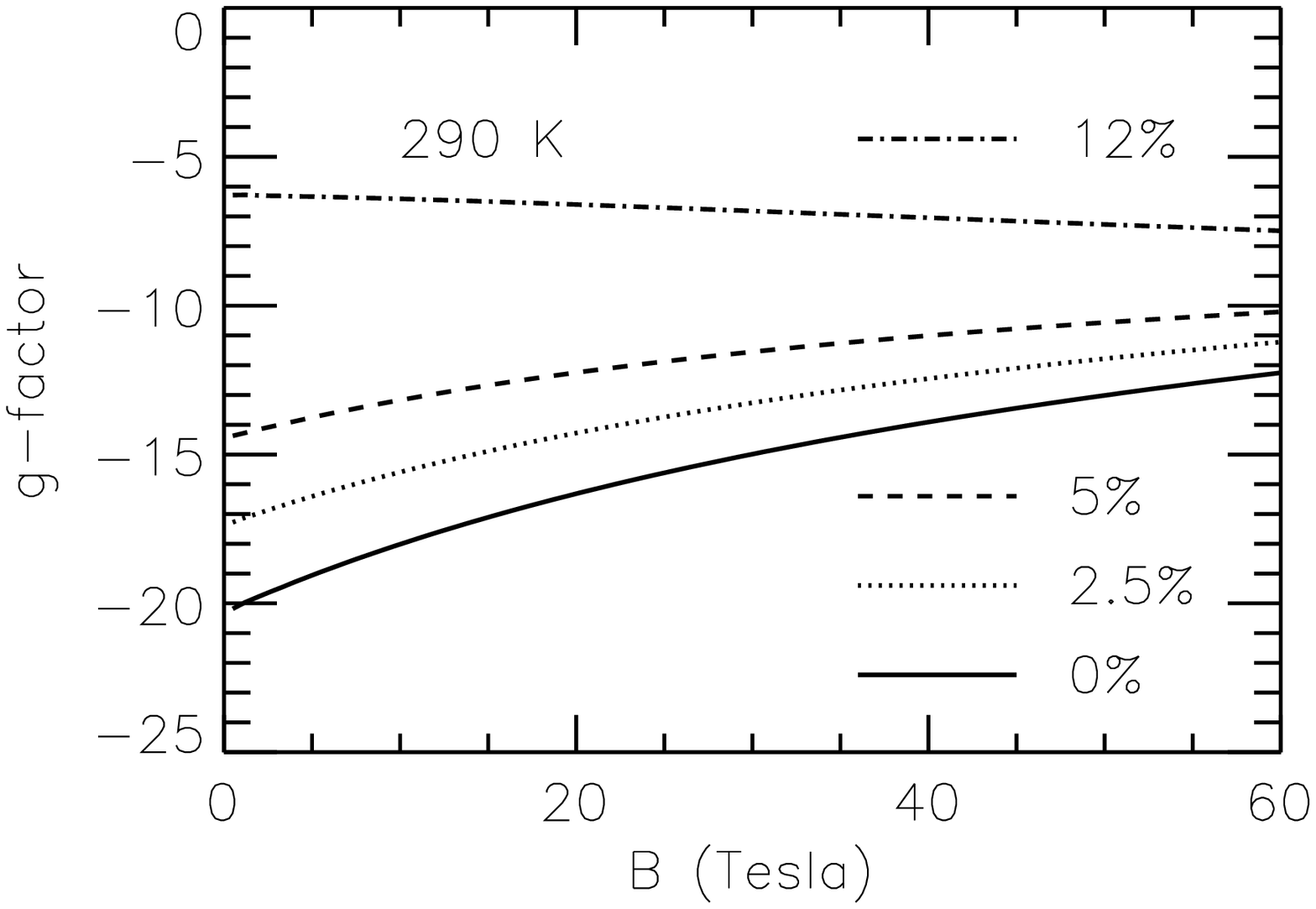}
\includegraphics[scale=0.48]{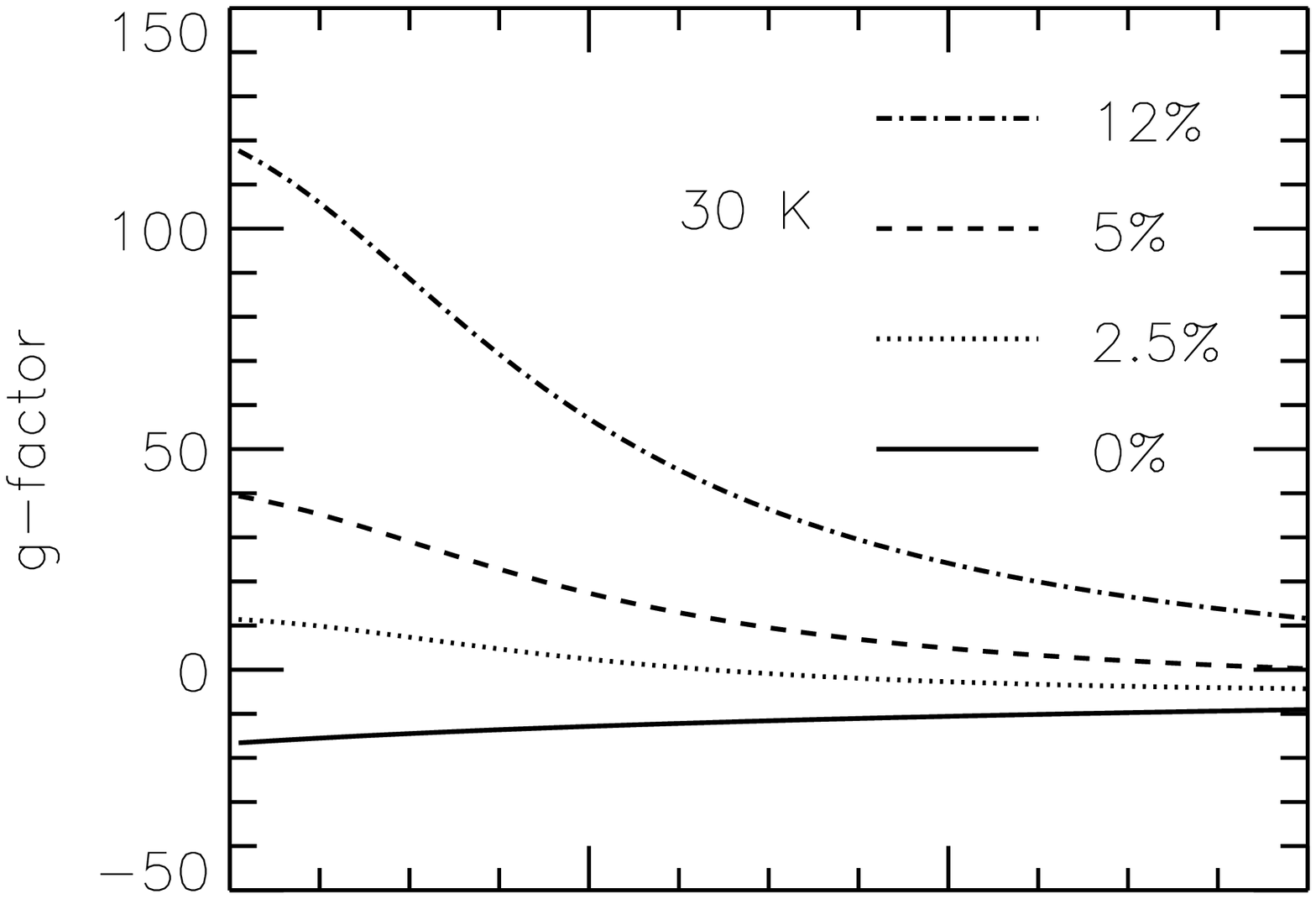}
\caption{\label{gbeps}
Gyromagnetic factors at $k=0$ for the lowest Landau subband in $n$-type
In$_{1-x}$Mn$_{x}$As as a function of applied magnetic field for several
values of the Mn concentration, x. The upper panel is for
$T = 30\ \mbox{K}$ and the lower panel is for $T = 290\ \mbox{K}$.}
\end{figure}

We have examined the electron g-factors for the lowest spin down
and spin up Landau levels with energies $E_{0,4}(k_z)$ and
$E_{1,6}(k_z)$, respectively. The gyromagnetic factor is obtained
by diagonalizing the Hamiltonians $H_0$ and $H_1$ in the
matrix eigenvalue problem (\ref{Schrodinger}) and computing
the gyromagnetic factor as
\begin{equation}
g = \frac{E_{0,4}(k_z)-E_{1,6}(k_z)}{\mu_B B}.
\end{equation}
The gyromagnetic factor can depend on the temperature, the magnetic
field, and the wavevector.

In Fig. \ref{gbeps} the gyromagnetic factor for the zone center is
plotted as a function of the magnetic field for two different
temperatures and four different values of the Mn concentration.
In all cases, doping with Mn impurities is seen to increase the
g factor and the effect is seen to be sensitive to both the
temperature and the Mn concentration. The sensitivity to temperature
and Mn concentration arises from the factor $x \ \langle S_z \rangle$
in the Mn exchange Hamiltonian (\ref{H_Mn}).
In the paramagnetic phase, $\langle S_z \rangle$ is related to the
temperature and magnetic field in accordance with
the simple Brillouin function expression of Eq. (\ref{S_z}).

\begin{figure}[tbp]
\includegraphics[scale=0.48]{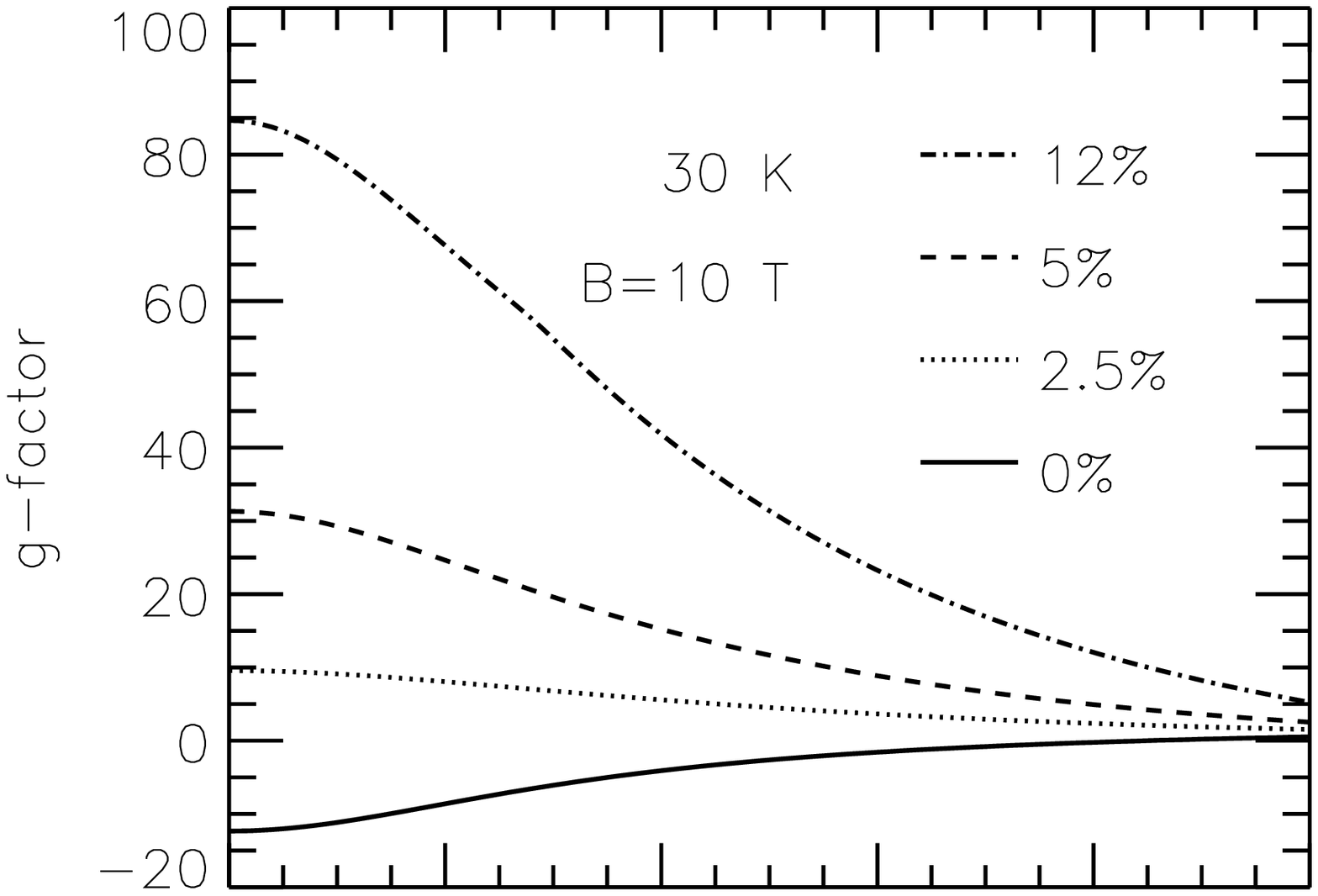}
\includegraphics[scale=0.48]{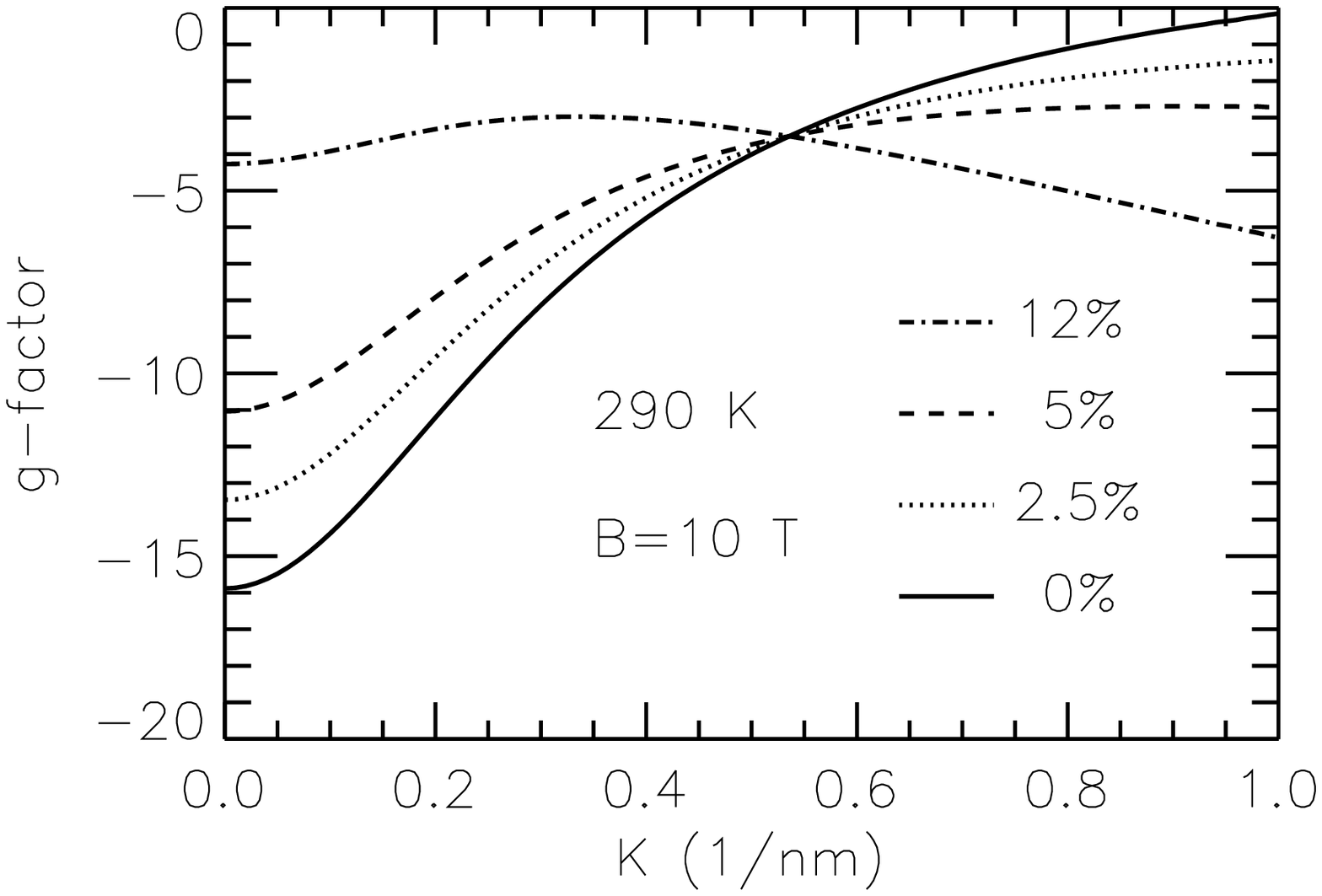}
\caption{\label{gkzeps}
Gyromagnetic factors at $B = 10\ \mbox{Tesla}$ for the lowest Landau
subband in $n$=type In$_{1-x}$Mn$_{x}$As as a function of $k$. $g$
factors for several values of Mn concentration are shown. The upper
panel is for $T = 30\ \mbox{K}$ and the lower panel is for
$T = 290\ \mbox{K}$.}
\end{figure}

In Fig. \ref{gkzeps}, the gyromagnetic factor is plotted as a
function of the wavevector $k_z$ for a magnetic field of $B$ = 10
T. As in Fig. \ref{gbeps}, two temperatures and four Mn
concentrations are considered. We note that in Fig.~\ref{gkzeps}(b) all
the curves cross around $k_z \approx 0.5 \ \mbox{nm}^{-1}$ indicating
that the exchange interaction contribution to the $g$ factor vanishes
for this state regardless of the Mn concentration. The reason for this
curious situation was explained earlier in our discussion of the Kane
model.

\subsection{Magneto-absorption}

\begin{figure}[tbp]
\includegraphics[scale=0.48]{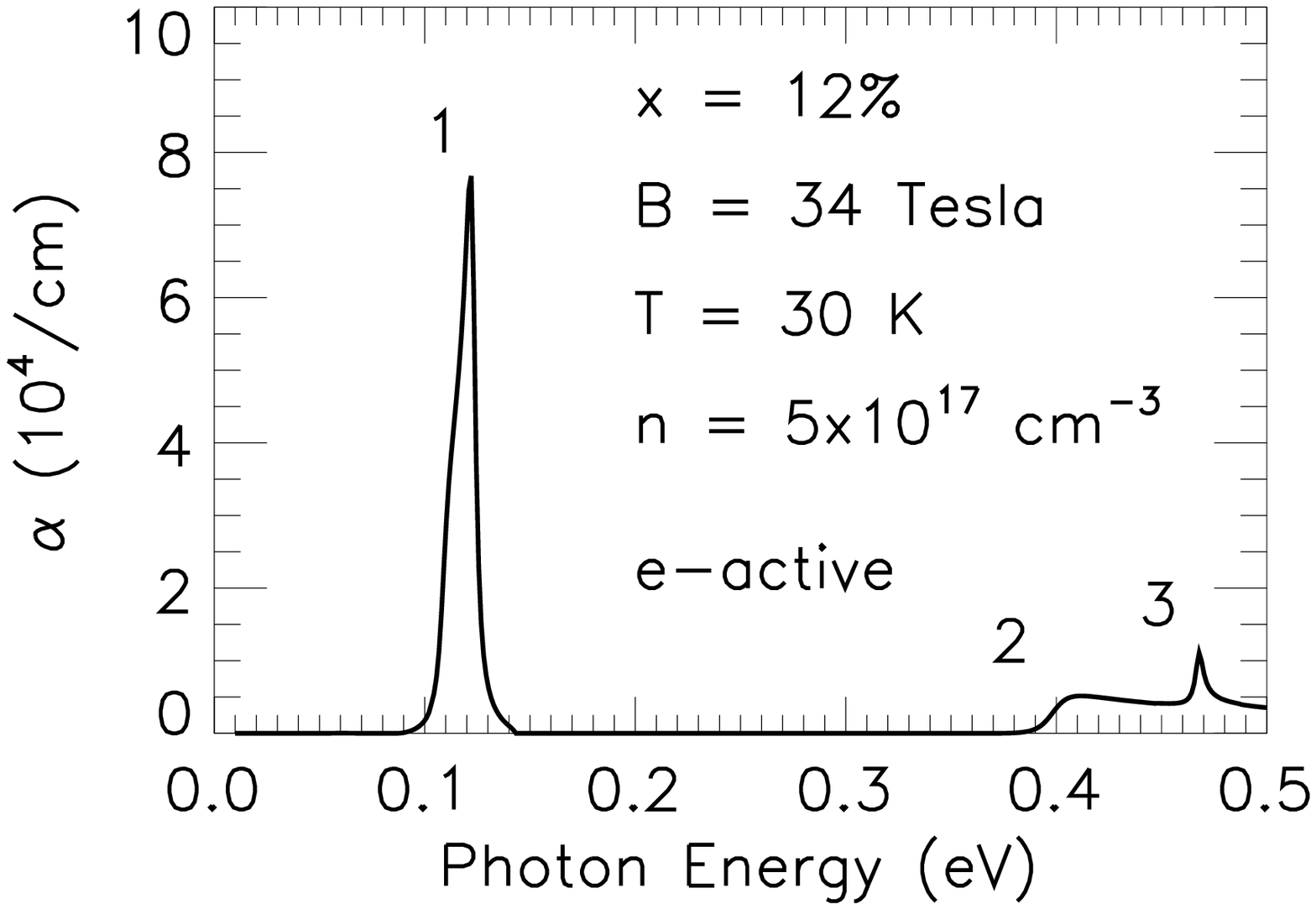}
\includegraphics[scale=0.48]{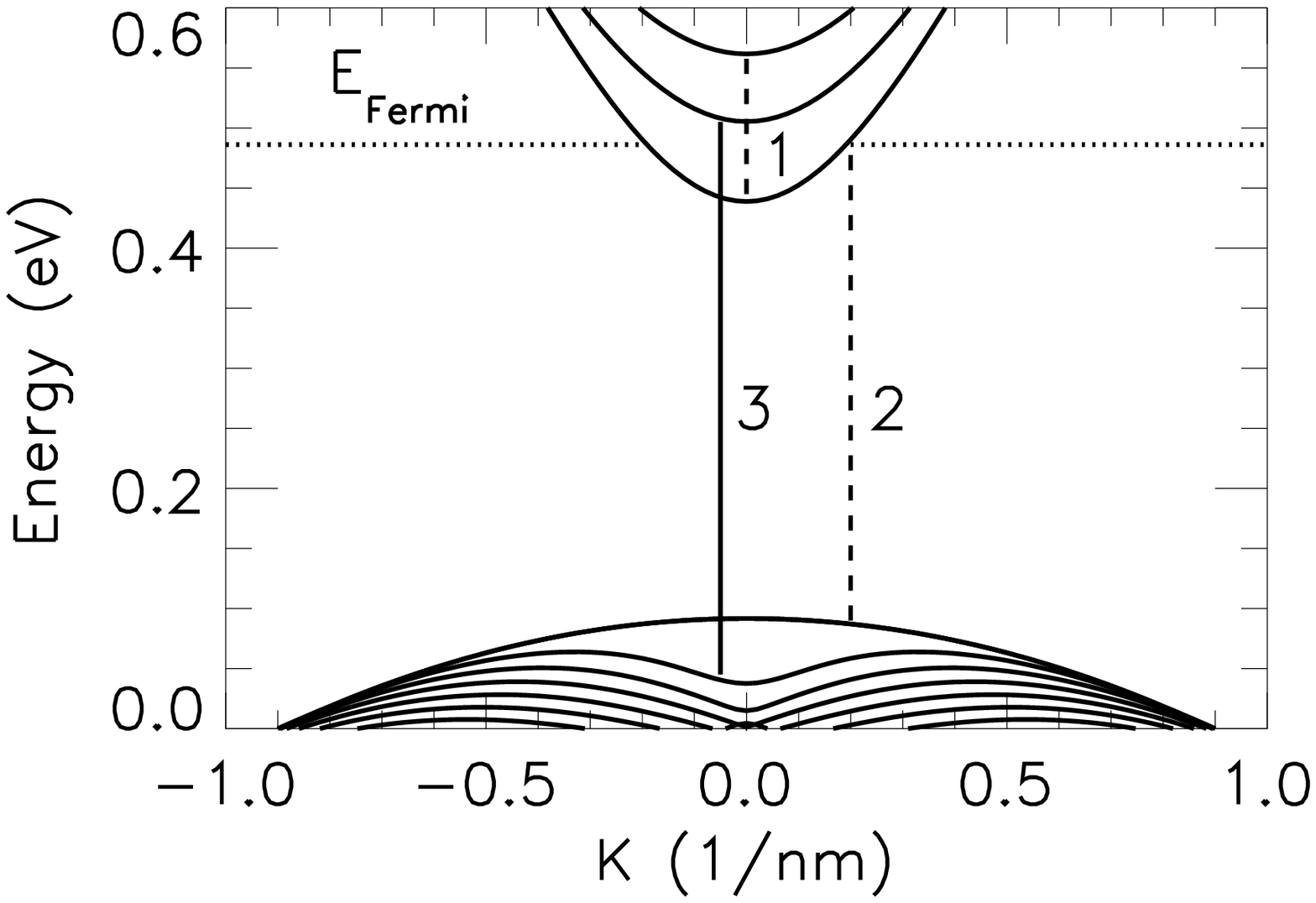}
\caption{\label{fig5} The upper panel shows the absorption
spectrum for $n$-doped In$_{0.88}$Mn$_{0.12}$As in a magnetic
field. The radiation is e-circularly polarized and the FWHM is
taken to be 4 meV. The bandstructure and Fermi energy are shown in
the lower panel. Three vertical transitions labelled 1, 2, and 3
correspond to absorption features in the upper panel. Transitions
between spin up levels are shown as solid lines and transitions
between spin down levels are shown as dashed lines.}
\end{figure}

The magneto-absorption spectrum and the bandstructure for
$n$-doped In$_{1-x}$Mn$_{x}$As are shown in the upper and lower
panels of Fig. \ref{fig5} for e-active circular polarization. In
our simulation, the Mn concentration is $x=12\%$, the external
magnetic field $B$ = 34 T, the carrier concentration $n=5 \times
10^{17}\ \mbox{cm}^{-3}$, and the temperature $T = 30\ \mbox{K}$.
The full width at half maximum (FWHM) is taken to be 4 meV in our
calculation. In the bottom panel of Fig. \ref{fig5} the Landau
subbands are plotted as a function of wavevector, $k_z$, parallel
to the applied magnetic field. The Landau subband energies depend
only on $k_z$ and the bandstructure is one dimensional. The Fermi
energy is indicated by the dotted line and at this carrier
concentration and temperature, only the lowest lying spin-down
conduction subband is populated near the zone center. Electrons in
this partially filled subband can be excited to higher lying
conduction subbands. A strong $\Delta n=1$ transition (labelled 1)
is observed between the filled ground state conduction subband and
the first excited spin-down conduction subband. Since these two
Landau levels have the same curvature, a sharply peaked joint
density of states results in the sharp peak in magneto-absorption
observed at a photon energy $\hbar \omega = 0.117\ \mbox{eV}$. Two
valence to conduction absorption features (labelled 2 and 3) are
seen for $\hbar \omega < 0.5\ \mbox{eV}$. Transitions between the
ground state spin-down valence and conduction subbands give rise
to the feature labelled 2 in the figure. The bandedge for this
absorption feature depends on the position of the Fermi energy due
to Fermi blocking effects. Another valence to conduction subband
transition (labelled 3) is due to transitions between the ground
state spin-up valence and conduction subbands. The valence subband
has a characteristic camel back structure and near the zone center
has the same curvature as the conduction subband. This results in
an enhancement in the joint density of states near the zone center
and gives rise to the peak in the absorption spectrum near $\hbar
\omega = 0.47\ \mbox{eV}$.

\subsection{Cyclotron resonance}

\begin{figure}[tbp]
\includegraphics[scale=0.48]{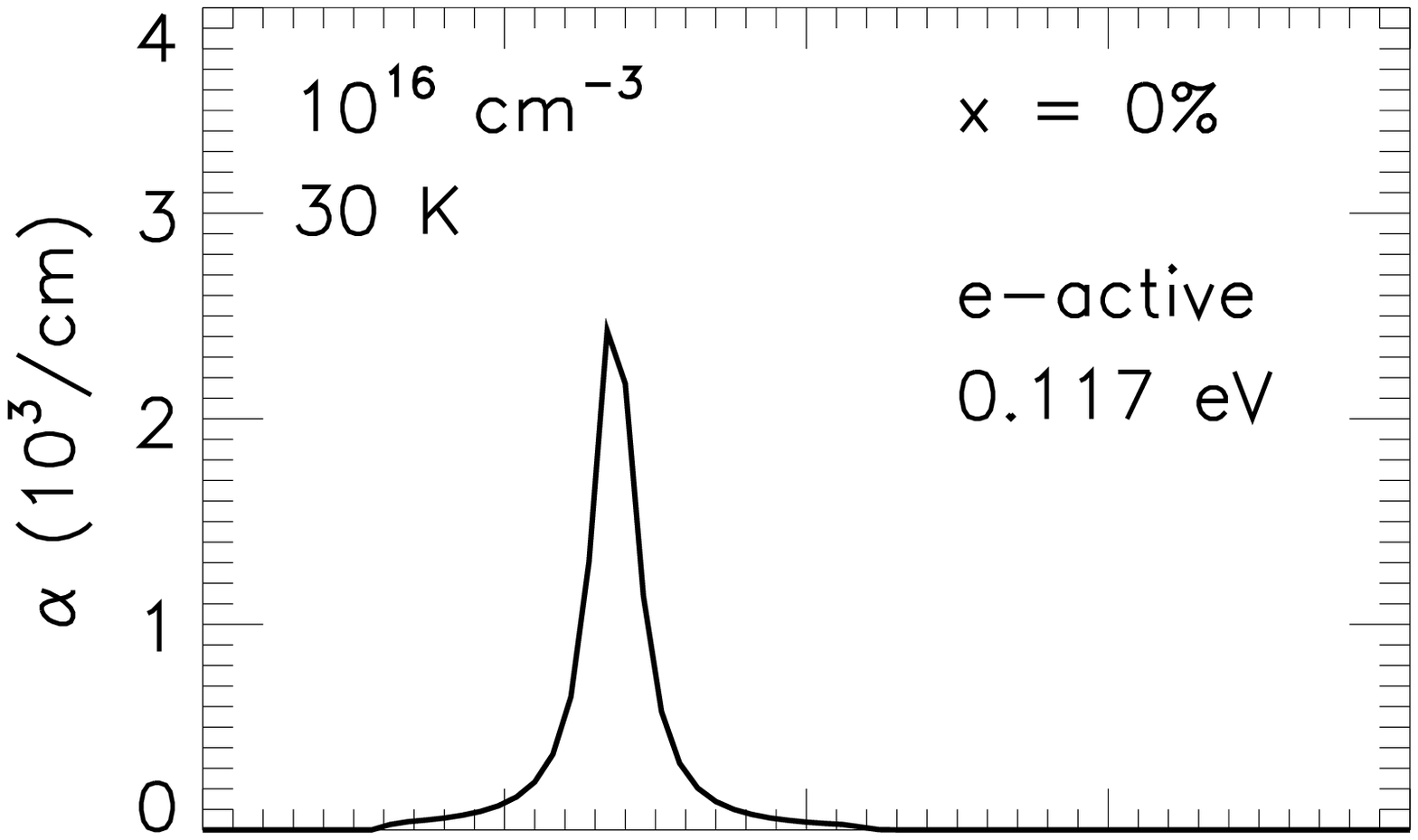}
\includegraphics[scale=0.48]{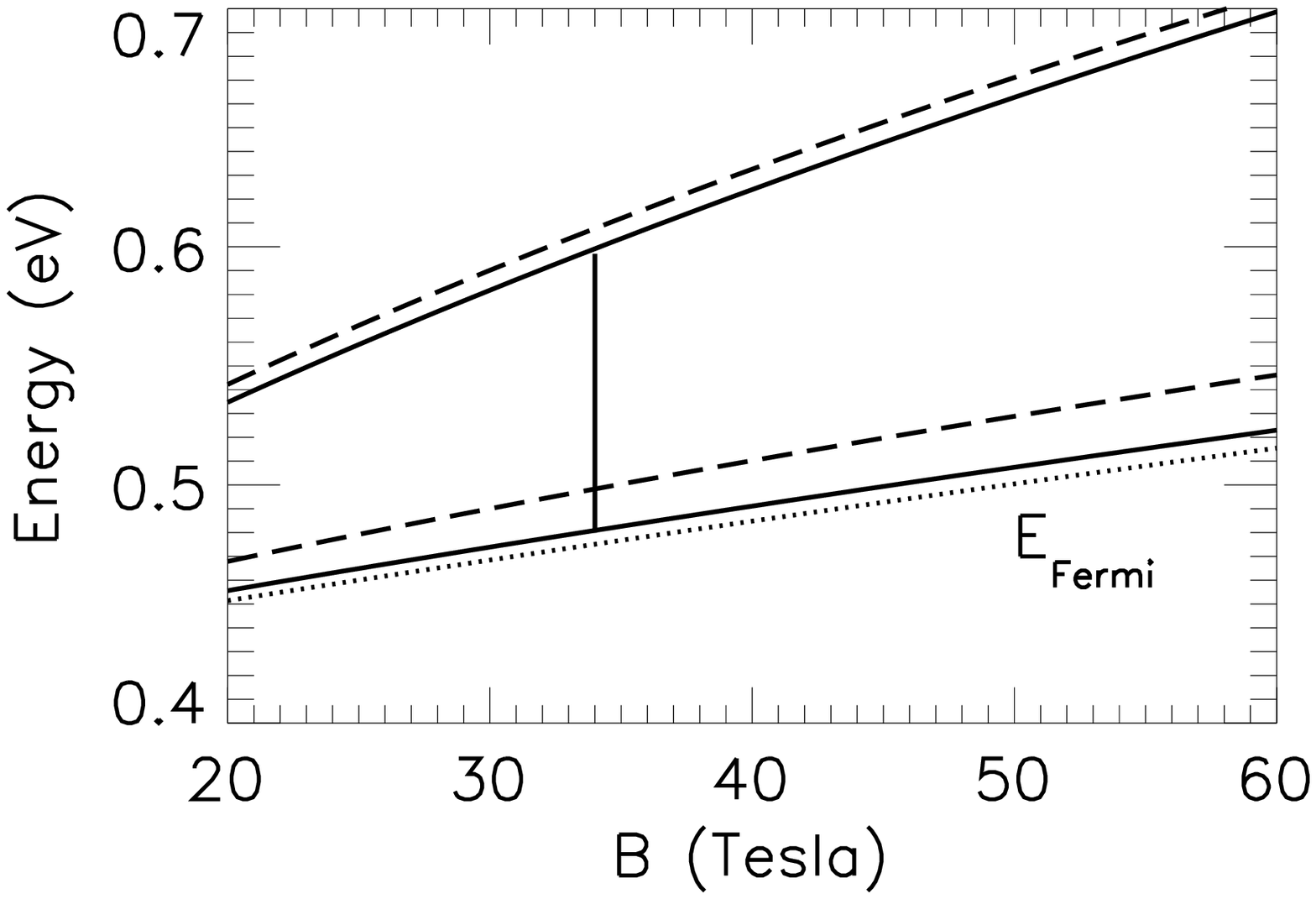}
\caption{\label{fig10}
The upper panel shows the cyclotron absorption as
a function of magnetic field in $n$-type InAs. The
radiation is e-active circularly polarized with
$\hbar \omega = 0.117 \ \mbox{eV}$. The FWHM linewidth is taken to
be 4 meV. The lower panel shows the
four lowest conduction Landau levels and the Fermi energy as a
function of applied magnetic field. Solid lines are spin-up levels
and dashed lines are spin-down.}
\end{figure}

In Fig. \ref{fig10}, we simulate cyclotron resonance experiments in
$n$-type InAs for e-active circularly polarized light with photon
energy $\hbar \omega=0.117\ \mbox{eV}$. We assume a temperature
$T = 30\ \mbox{K}$ and a carrier concentration
$n=10^{16}\ \mbox{cm}^{-3}$.
The lower panel of Fig. \ref{fig10} shows the four lowest zone center
Landau conduction subband energies and the Fermi energy as functions
of the applied magnetic field. The transition at the resonance energy
$\hbar \omega=0.117\ \mbox{eV}$ is a spin-up $\Delta n=1$
transition and is indicated by the vertical line. From the Landau
level diagram the resonance magnetic field is found to be
$B$ = 34 T. The upper panel of Fig. \ref{fig10} shows
the resulting cyclotron resonance absorption assuming a FWHM linewidth
of 4 meV. There is only one resonance line in the cyclotron absorption
because only the ground state Landau level is occupied at low electron
densities.

\begin{figure}[tbp]
\includegraphics[scale=0.48]{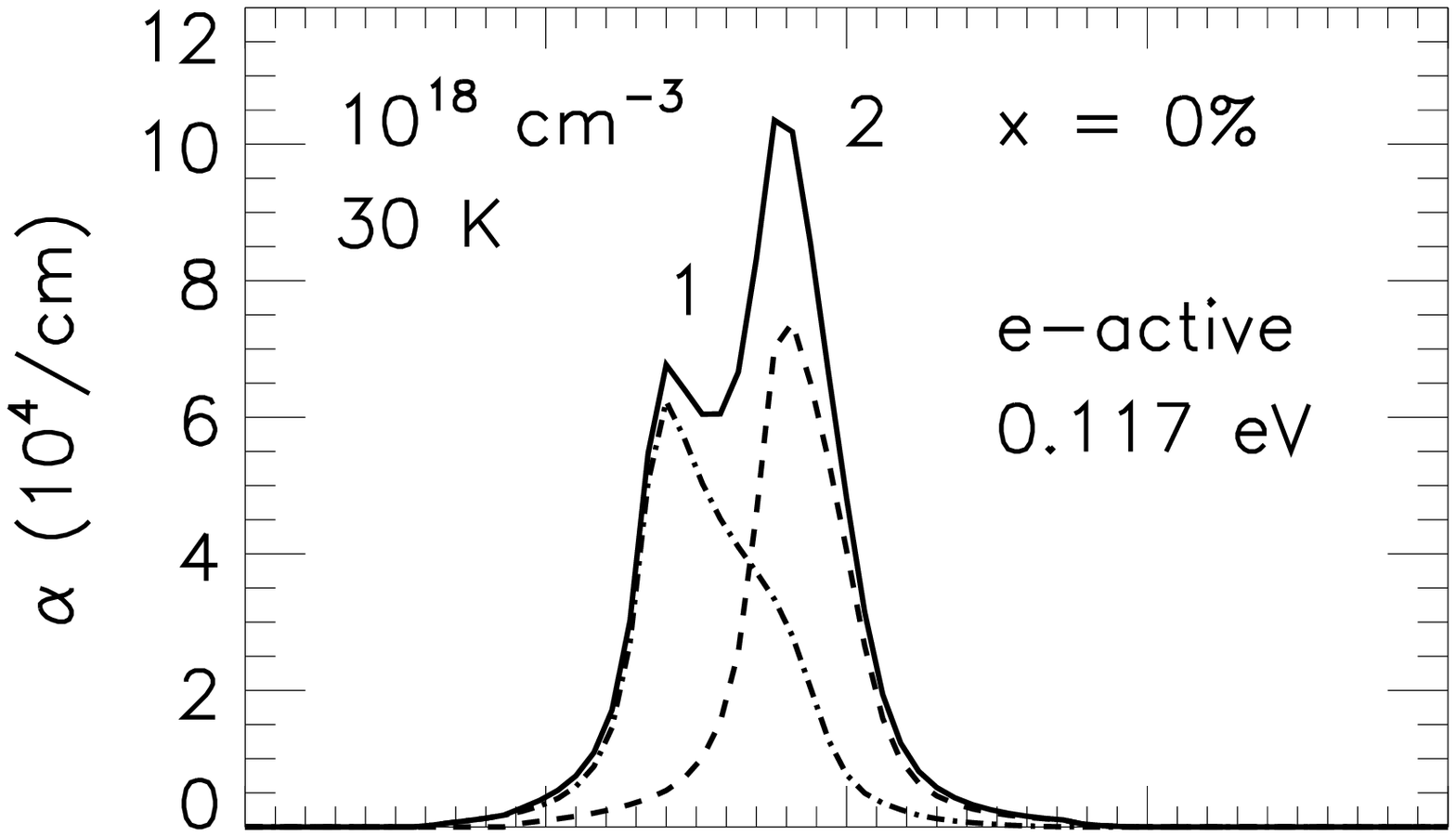}
\includegraphics[scale=0.48]{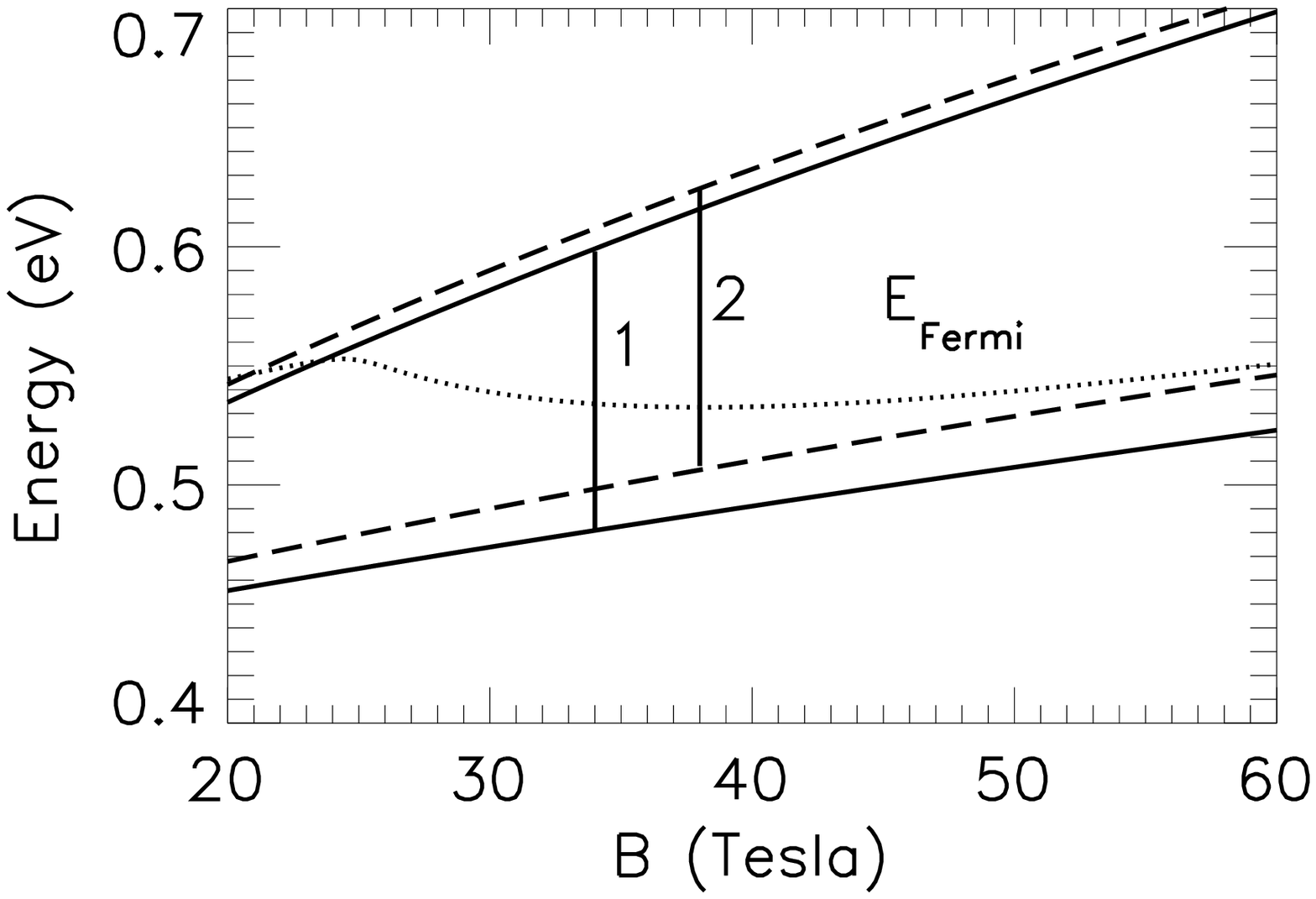}
\caption{\label{fig7}
Same simulation as in Fig. \ref{fig10} but with the doping
density raised to $n = 10^{18} \mbox{cm}^{-3}$. Since the two
lowest Landau levels are occupied, there are now two cyclotron
absorption peaks. Peak 1 is the spin up transition and Peak 2
is an additional spin down transition. The dashed-dotted and dashed
lines show the individual contributions to the cyclotron resonance
absorption 
from transition 1 and 2 respectively.  The asymmetry of the lineshape
results from the $k_z \ne 0$ contributions as well as the
nonparabolicity.}
\end{figure}

At higher electron densities multiple lines can appear in the cyclotron
absorption as higher lying Landau subbands are populated. This is the
spin-splitting of the cyclotron resonance peaks.  This is illustrated
in Fig. \ref{fig7} where the electron density, $n$, has been increased
to $10^{18}\ \mbox{cm}^{-3}$. We can see in Fig. \ref{fig7} that both
the lowest lying spin-up and spin-down Landau levels are occupied and
that two resonance transitions (labelled 1 and 2 in Fig.  \ref{fig7})
result. In addition, we also see that the spin-down peak which comes in
at higher electron densities appears to dominate the spin-up peak.
This is somewhat misleading. The  dashed-dotted and dashed lines show
the contributions to the cylotron resonance absorption from transitions
1 and 2 respectively. Both transitions have roughly the same strength,
but the lineshape for transition 1 is highly asymmetric.  This results
from taking into account the contributions to the absorption for
$k_z \ne 0$ and also the nonparabolicity of the bands. The lineshape for
transition 2 is not as assymetric since the Fermi energy lies closer to
the Landau level edge for this transition.  Note that if there is
substantial scattering so that the transition lines are broadended,
then the two transition will be merged into one and can lead to an
apparent  shift of cyclotron resonance features to higher fields.

\begin{figure}[tbp]
\includegraphics[scale=0.48]{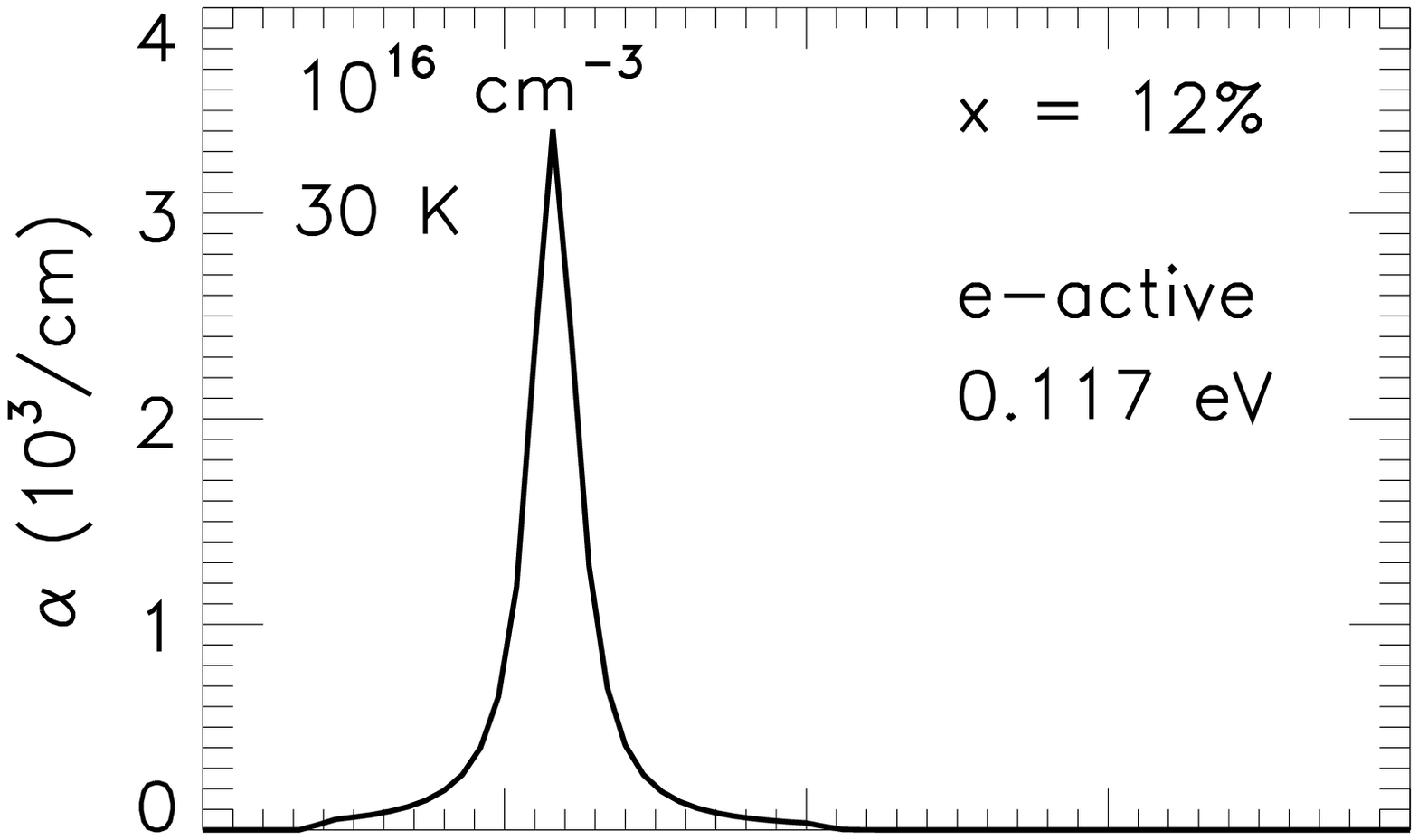}
\includegraphics[scale=0.48]{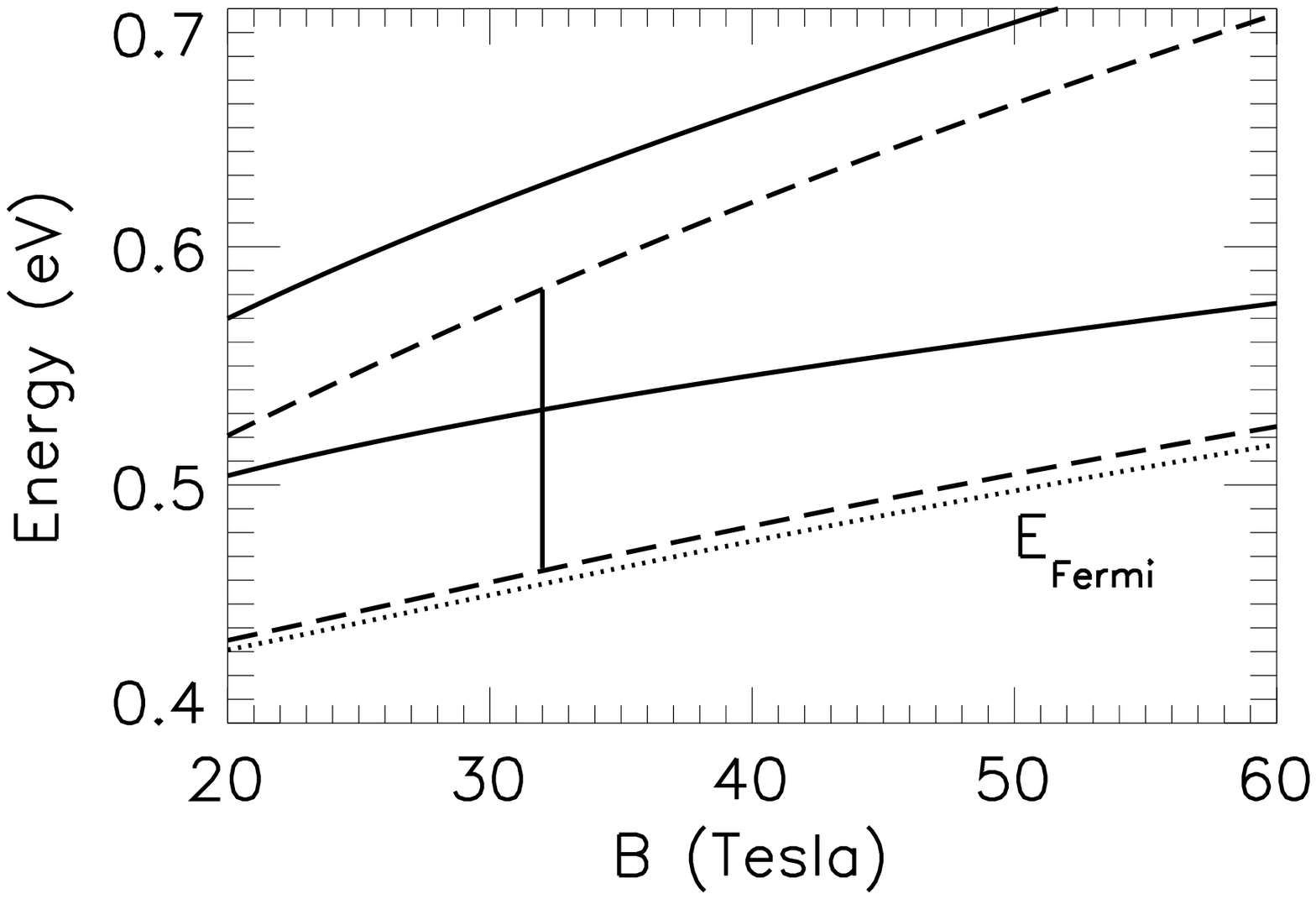}
\caption{\label{fig11}
Same simulation as in Fig. \ref{fig10} but with the
Manganese concentration increased to $x = 12\%$. The cyclotron
absorption peak near 32 T is now a spin down $\Delta n = 1$
transition.}
\end{figure}

The effect on the cyclotron resonance absorption of doping InAs
with $12\%$ Mn is shown in Fig. \ref{fig11}. Comparing Figs.
\ref{fig10} and \ref{fig11}, we see that the effect of heavily
doping with Mn while keeping the electron concentration, $n$,
fixed is to reverse the spin splitting and hence the character and
position of the cyclotron absorption peak. The cyclotron
absorption peak seen in Fig. \ref{fig11} is a spin-down transition
as opposed to a spin-up transition and the cyclotron resonance
peak is seen to shift from $B$ = 34 T to $B$ = 32 T.

\begin{figure}[tbp]
\includegraphics[scale=0.48]{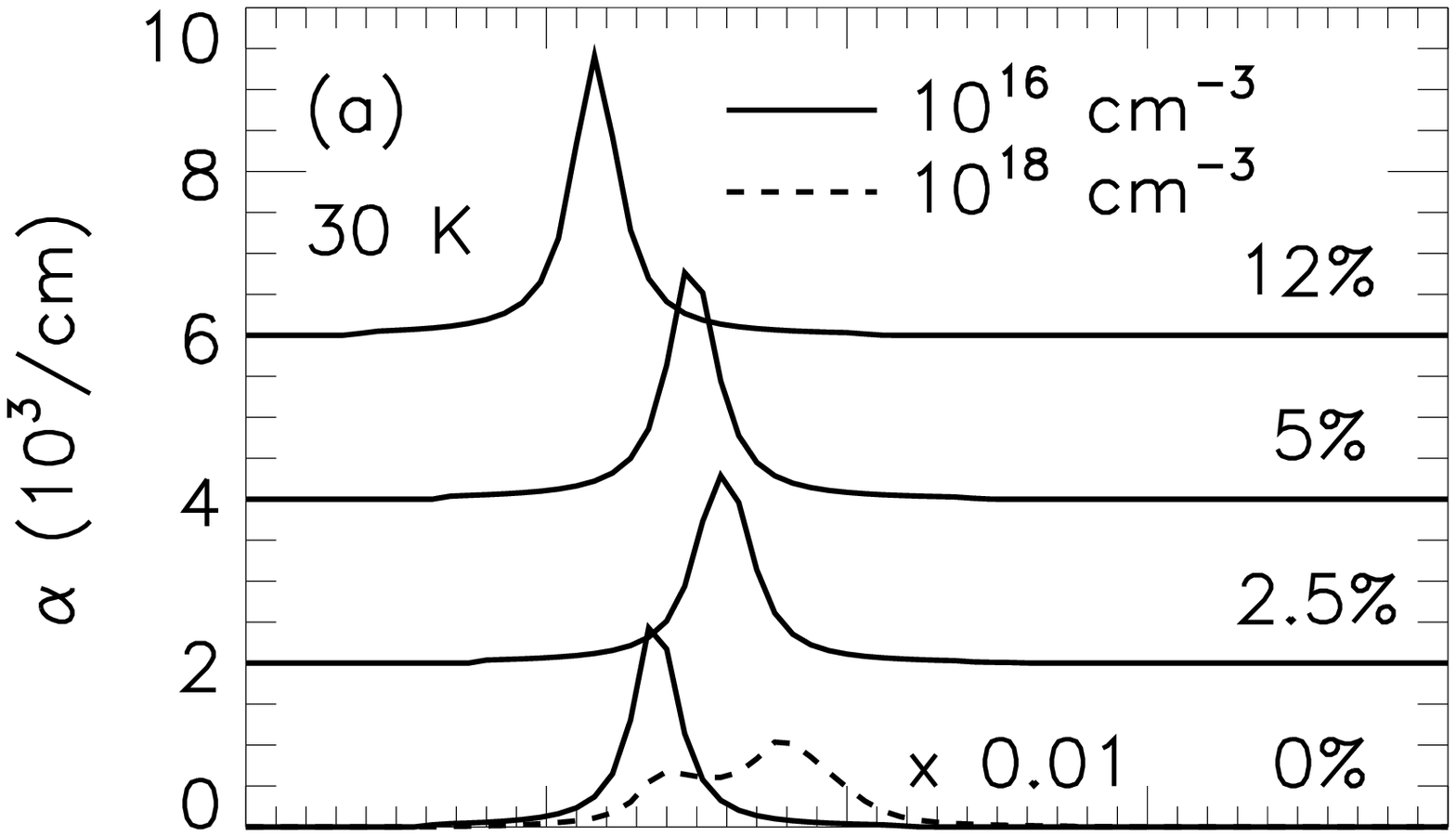}
\includegraphics[scale=0.48]{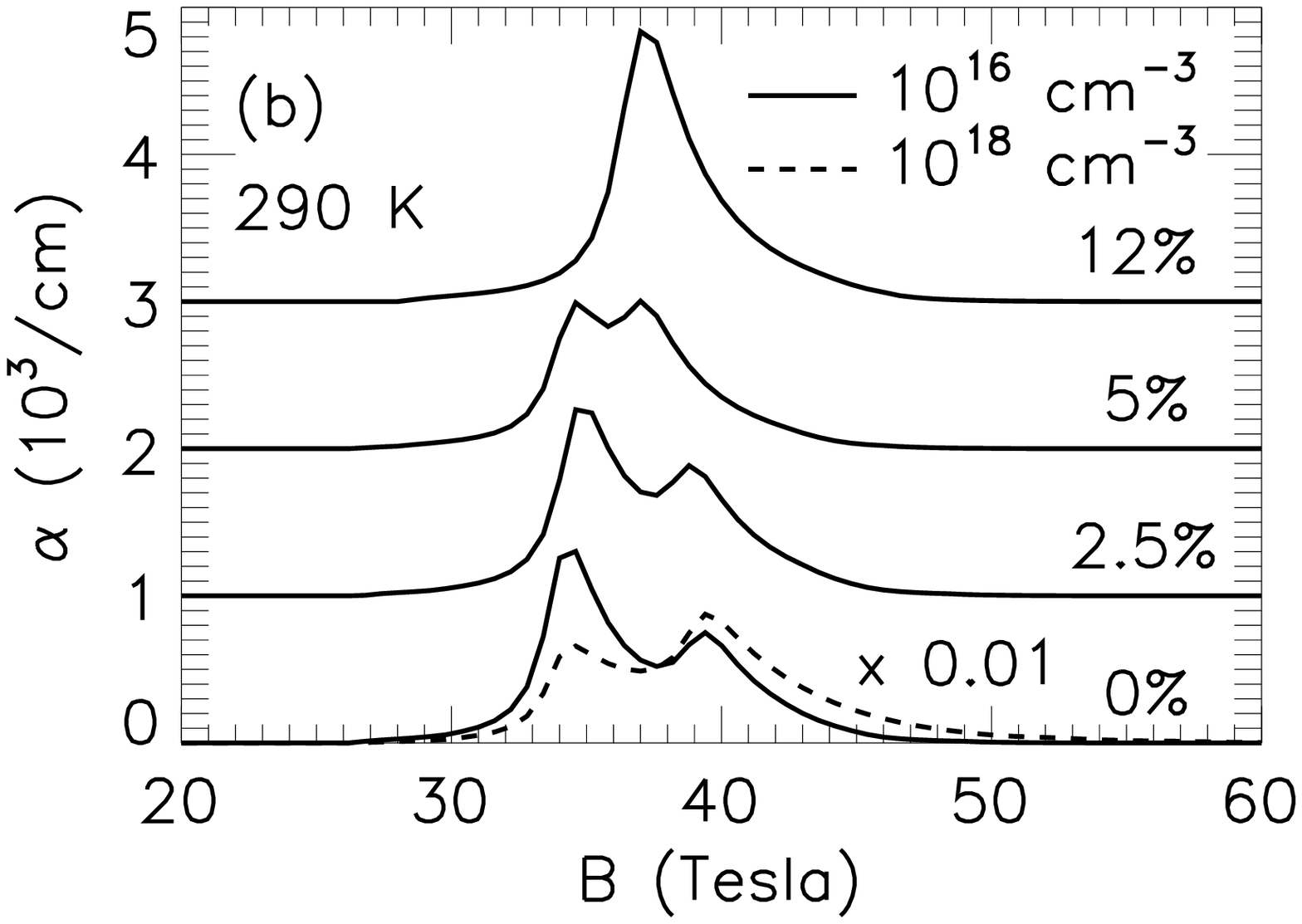}
\caption{\label{fig1}
Cyclotron absorption as a function
of magnetic field in $n$-doped In$_{1-x}$Mn$_{x}$As with
$x=0\%$, $2.5\%$, $5\%$, and $12\%$ for (a) $T = 30 \ \mbox{K}$
and (b) $T = 290 \ \mbox{K}$. The radiation is e-active
circularly polarized with $\hbar \omega = 0.117 \ \mbox{eV}$.}
\end{figure}

In Fig. \ref{fig1}, we examine more closely the effects of the Mn
doping concentration, $x$, on the cyclotron absorption spectra of
$n$-doped In$_{1-x}$Mn$_{x}$As for e-active circularly polarized
radiation with photon energy $\hbar \omega=0.117\ \mbox{eV}$. We
plot the cyclotron absorption as a function of applied magnetic
field for $x=0\%$, $2.5\%$, $5\%$, and $12\%$. In all cases we
assume a fairly narrow FWHM linewidth of 4 meV. In Fig. \ref{fig1}
(a) we plot the cyclotron absorption at $T = 30\ \mbox{K}$. All
the solid curves are computed for an electron concentration
$n=10^{16}\ \mbox{cm}^{-3}$ and are vertically offset for clarity.
For $x=0\%$, we plot the cyclotron absorption for $n=10^{18}\
\mbox{cm}^{-3}$ as a dashed line scaled by a factor of 0.01. In
Fig. \ref{fig1} (a), the cyclotron absorption curves for $x=0\%$
and $x=12\%$ have already been discussed in Figs. \ref{fig10},
\ref{fig7} and \ref{fig11}. As $x$ increases from $0\%$ to $12\%$,
the cyclotron absorption peak initially shifts to higher fields
and is due to a $\Delta n = 1$ spin-up transition from the
occupied lowest Landau level. At a critical value of the Mn
concentration, the spin splittings reverse and the observed
transition is now due to a $\Delta n = 1$ spin-down transition.
Consequently, the cyclotron absorption begins to decrease with
increasing Mn concentration.

To examine the temperature dependence, we also computed the cyclotron
absorption at room temperature for several values of $x$.
The curves in Fig. \ref{fig1} (b) are the same as those in
Fig. \ref{fig1} (a) except that the temperature has been increased
from $T = 30\ \mbox{K}$ to $T = 290\ \mbox{K}$.
We see that the single cyclotron absorption peaks observed
at low electron density in Fig. \ref{fig1} (a) split into doublets
in Fig. \ref{fig1}(b). At room temperature, the two lowest Landau
levels are thermally populated and both the spin-up and
spin-down transitions appear in the cyclotron absorption spectra.
The shift in the cyclotron absorption features with increasing
Mn concentration is also different at room temperature.
As $x$ increases, we find that the cyclotron absorption features
in Fig. \ref{fig1} (b) shift to higher fields. As we increase the
temperature the magnitude of the
average Mn spin $\langle S_z \rangle$ described by
the Brillouin function in Eq. (\ref{S_z}) decreases. Since the exchange
Hamiltonian, $H_{Mn}$, in Eq. (\ref{H_Mn}) is proportional to
$x\ \langle S_z \rangle$, the spin splittings reverse sign at higher
Mn concentrations.

\subsection{Electron cyclotron mass}
\label{cyclotron-mass}
\begin{figure}[tbp]
\includegraphics[scale=0.48]{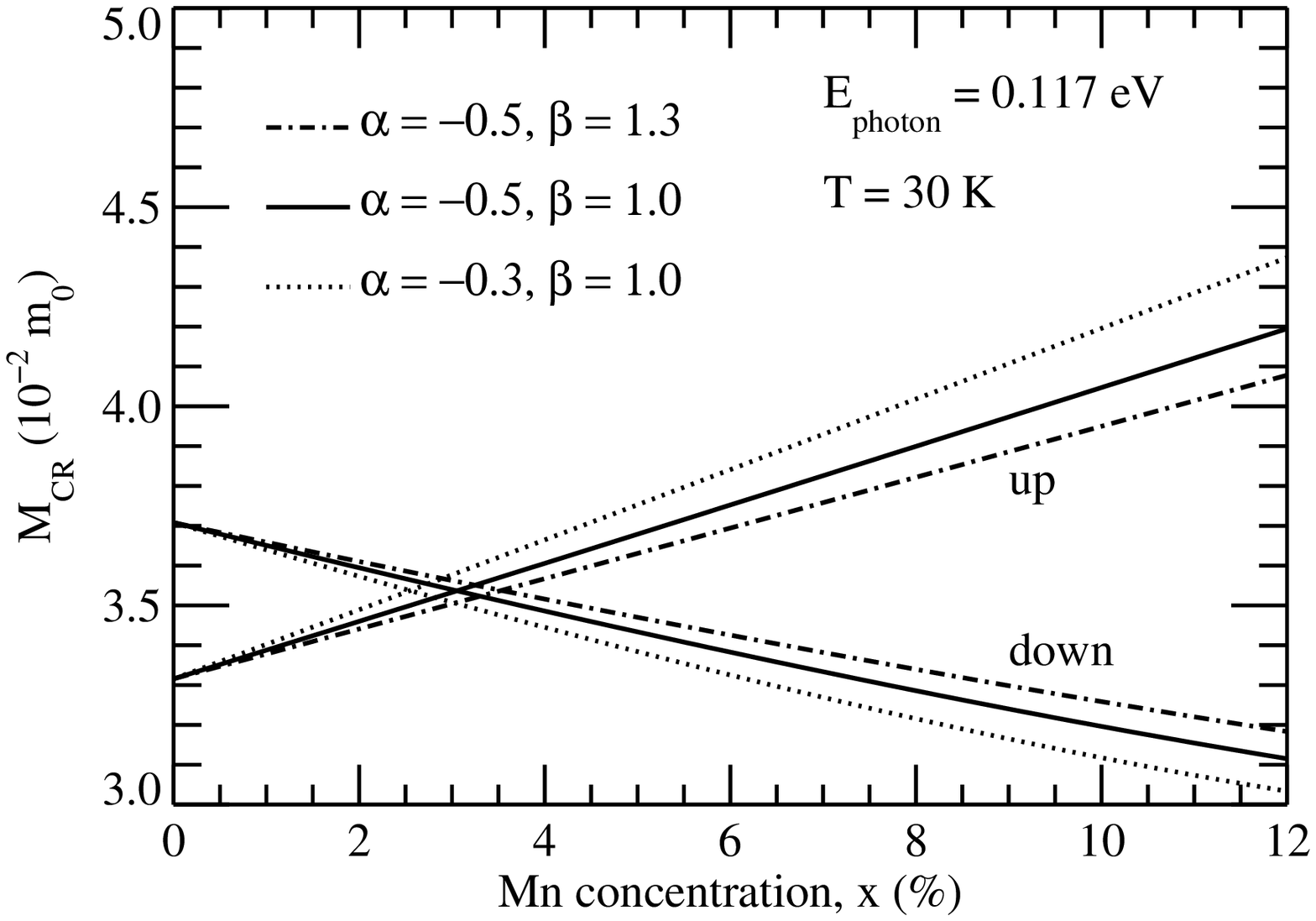}
\includegraphics[scale=0.48]{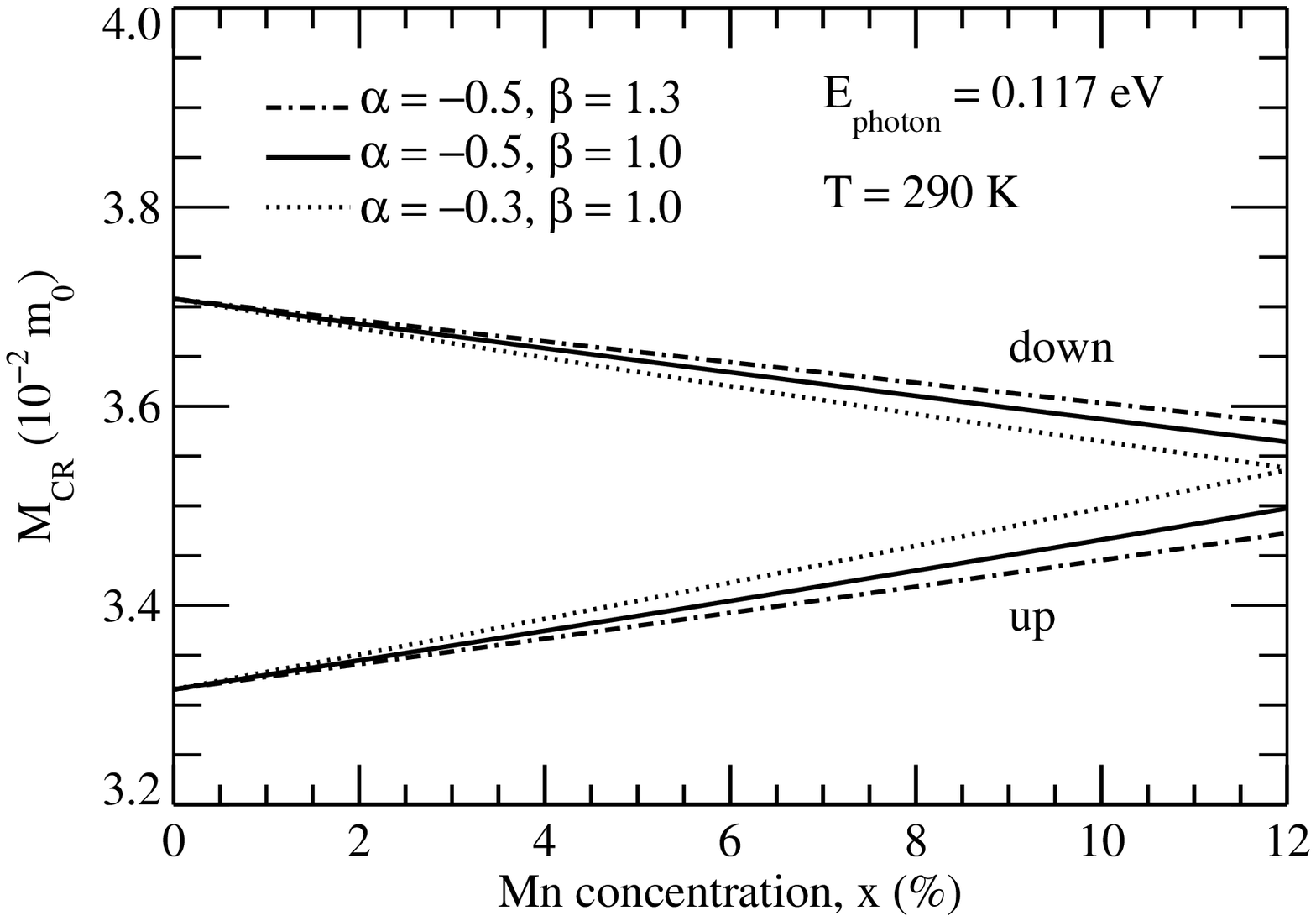}
\caption{\label{fig15}
Calculated electron cyclotron masses for the lowest lying spin up
and spin down Landau transitions in $n$-type In$_{1-x}$Mn$_{x}$As
at $\hbar \omega = 0.117 \ \mbox{eV}$ as a function of Manganese
concentration for (a) $T = 30 \ \mbox{K}$ and
(b) $T = 290 \ \mbox{K}$. Electron cyclotron masses
are shown for three sets of $\alpha$ and $\beta$ values.}
\end{figure}

The electron cyclotron mass, $M_{CR}$, for a given cyclotron absorption
transition is related to the resonance field, $B^{*}$
(cyclotron energy $\mu_B B^{*}$), and photon
energy, $\hbar \omega$, by the definition
\begin{equation}
\frac{M_{CR}}{m_0} \equiv \frac{2 \mu_B B^{*}}{\hbar \omega}
\label{M_CR}
\end{equation}
In Fig. \ref{fig15}, the calculated cyclotron masses for the lowest
spin-down (b set) transitions and spin-up (a set) transitions are
plotted as a function of Mn concentration, $x$, at a photon energy
of $\hbar \omega = 0.117\ \mbox{eV}$. Cyclotron masses are computed
for several sets of $\alpha$ and $\beta$ values.
The cyclotron masses in
Fig. \ref{fig15} (a) and (b) correspond to the computed cyclotron
absorption spectra shown in Fig. \ref{fig1} (a) and (b), respectively.
In our model, the electron cyclotron masses shown depend on the
Landau subband energies and photon energies and are independent of
electron concentration (though clearly, the population of a given
state will depend upon the concentration).

In the Kane model discussed earlier, the cyclotron energy in
In$_{1-x}$Mn$_{x}$As depends linearly on the Mn concentration, $x$.
In addition, the cyclotron mass, which is
inversely proportional to the cyclotron
resonance energy, increases for conduction electrons from
the (a) subsystem and decreases for conduction electrons in the (b)
subsystem. These predictions of the simple Kane model are both
confirmed as can be seen in Fig. \ref{fig15}.

The sensitivity of the cyclotron masses to $\alpha$ and $\beta$ seen in
Fig.~\ref{fig15} can be understood if we study the dependence of the
cyclotron energy on the exchange constants $\alpha$ and $\beta$ in the
simple Kane model previously discussed.  Because InMnAs is a narrow
gap semiconductor, it is not surprising that the conduction band
cyclotron energy should depend not only on the conduction band exchange
constant $\alpha$ but also on the valence band exchange constant
$\beta$.  This just reflects conduction-valence band mixing in
the magnetic field. Our Kane model calculations show, however, that
only their difference is important and that only one independent
constant $(\alpha-\beta)$ is needed to describe conduction band
cyclotron resonance with high accuracy. It is seen from
Eqs.~(\ref{CEal})-(\ref{CEbh}) that in the high-field limit the
exchange interaction correction to the cyclotron energy is proportional
to $(\alpha-\beta)$ while in the low-field limit the term proportional
to $(\alpha-\beta)$ is one order of magnitude larger than the term
proportional to $\beta$. Most of the important corrections to the
cyclotron energy come from the heavy hole admixture to conduction band
states. The difference $(\alpha-\beta)$ reflects the change in the band
gap due to the exchange interaction  and affects the degree of this
mixing.  This suggests that the dependence on $(\alpha-\beta)$ can be
interpretted as just a renormalization of the band gap in a magnetic
field.

Note that while Fig.~\ref{fig15} shows that the cyclotron peak
positions are very sensitive to ($\alpha-\beta$) (the figure shows
how a 20\% change in ($\alpha-\beta$)
can affect these positions), one can only measure the
difference $\alpha - \beta$ from these measurements and not
$\alpha$ and $\beta$ independently. In addition, to determine
($\alpha-\beta$) the
other parameters (such as $\gamma_1$, $\gamma_2$, $\gamma_3$, etc.)
must be accurately known.

\subsection{Comparison with experiment}

In this section, we compare the results of our 
theoretical calculations with our experimental results.

\begin{figure}
\includegraphics[scale=0.48]{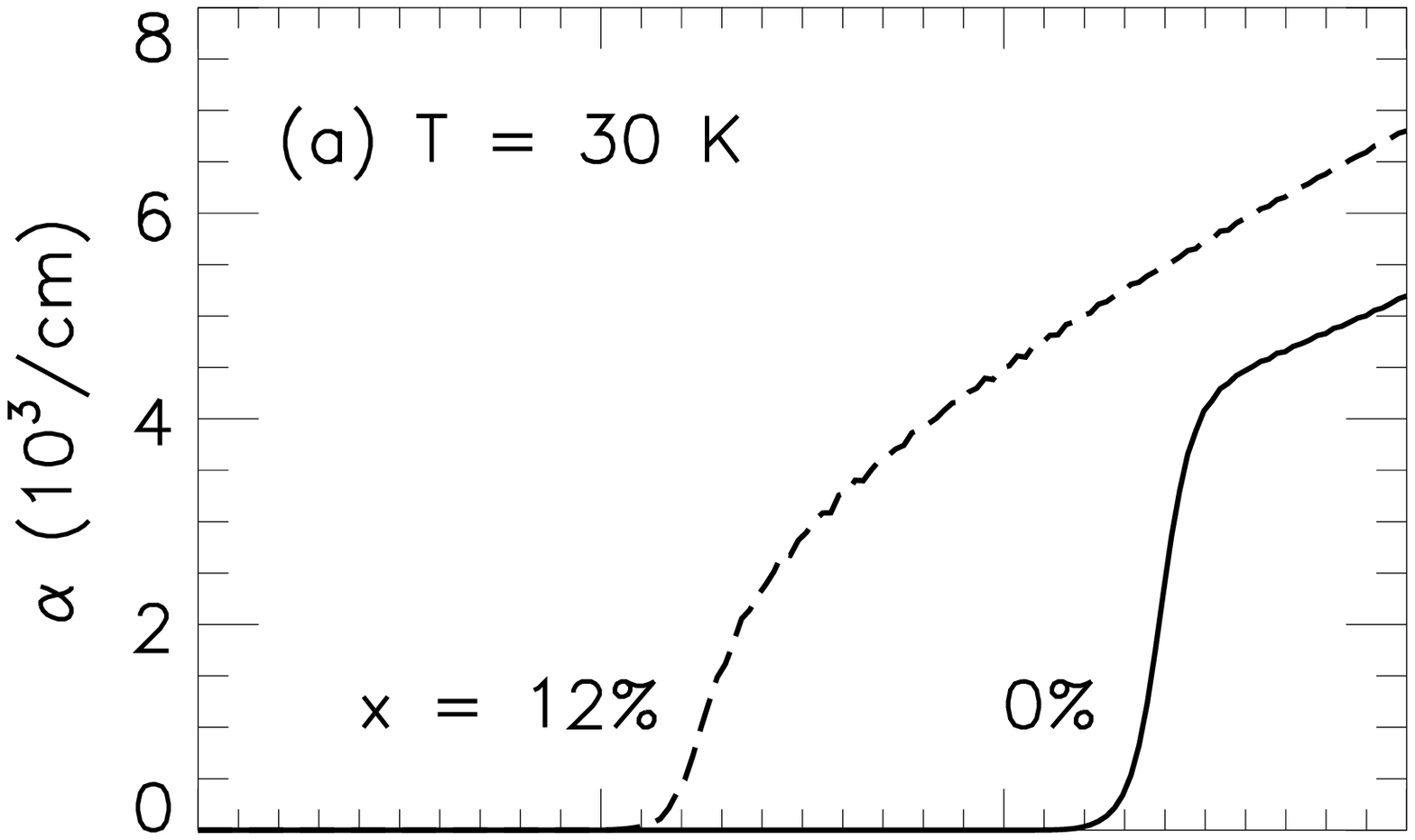}
\includegraphics[scale=0.48]{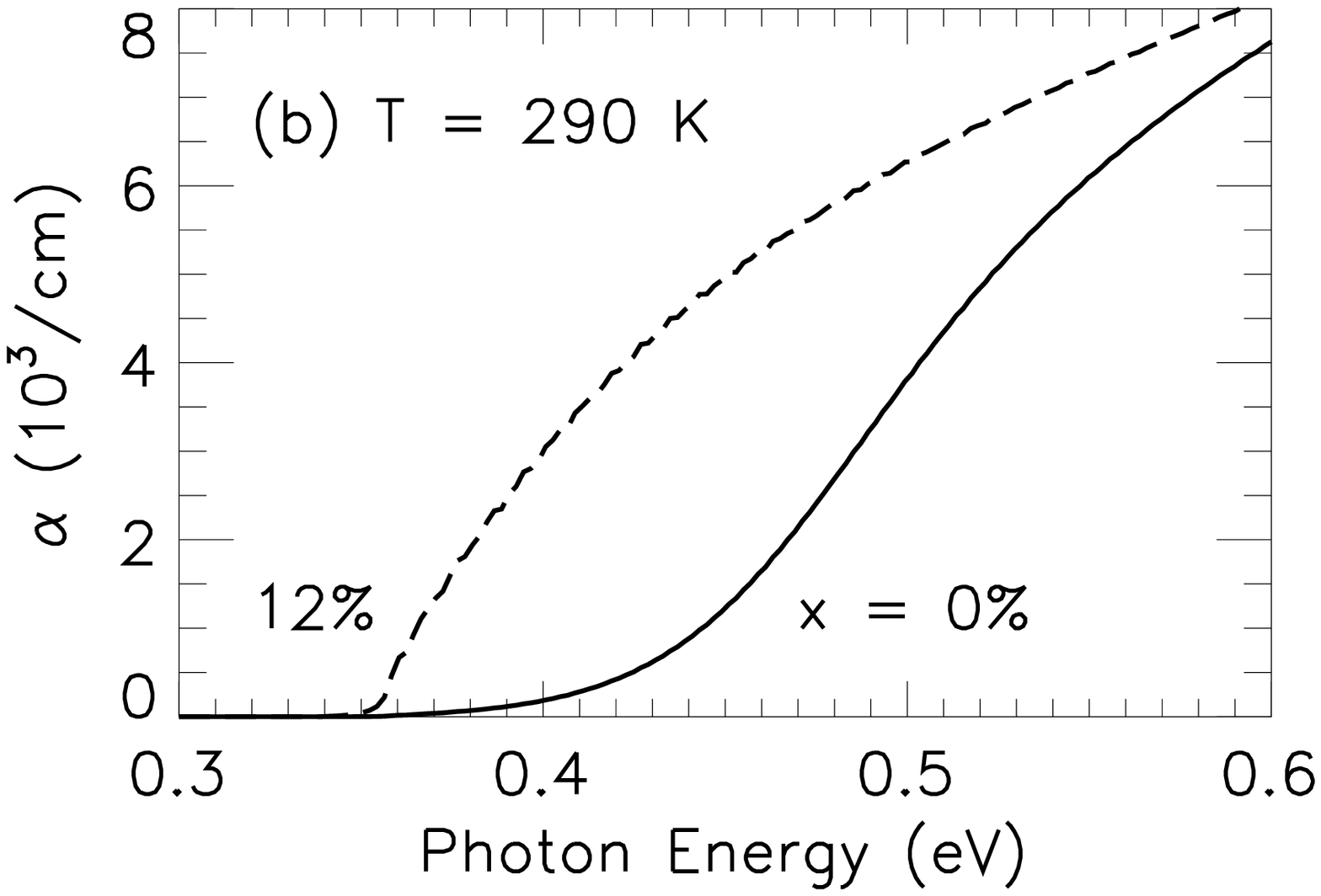}
\caption {Theoretical absorption coefficient for linear polarization
as a function of photon energy at 30 K (a) and 290 K (b) at $B = 0$
for Mn concentrations $x = 0\%$ (solid lines) and $x = 12\%$
(dashed lines).
Shifts in the bandgaps are entirely due to carrier filling effects.
These figures should be compared with the experimental data shown in
Fig. \ref{figftir}.}
\label{theory2}
\end{figure}

We first compute the near band gap absorption spectra for $x$ = 0 and
0.12.  This is important since in our theoretical calculations, we have
only included the effect of $sp-d$ interaction of the Mn ions with
the electrons and holes.  An additional, non-magnetic effect of the Mn
ions could be to change the band gap with doping $x$ similar to the
band gap shift of $Al_xGa_{1-x}As$ with increasing $Al$ content.
Figures \ref{theory2}(a) and \ref{theory2}(b) show calculated
near-bandgap absorption spectra for the $x$ = 0 and 0.12 samples at (a)
30 K and (b) 290 K without taking into account a change in band gap
with Mn doping.   The theory successfully reproduces the experimental
results shown in Fig.~\ref{figftir}.  The blue shift of the band gap in
the $x$ = 0 sample is entirely due to band filling effects, i.e., the
Burstein-Moss shift. The $x=0$ sample has a relatively high electron
density.  As Mn doping is increased, the electron density decreases
since the Mn ion acts as an acceptor.  The theory curves in the
figure also show the sharpening of the band edge at low temperatures,
consistent with the experiment.  The fact that our theory curves are
able to
reproduce the $x$ dependent absorption spectra without changing the band
gap indicates that any $x$ dependence to the band gap is small and we 
therefore neglect it.

\begin{figure}
\includegraphics[scale=0.5]{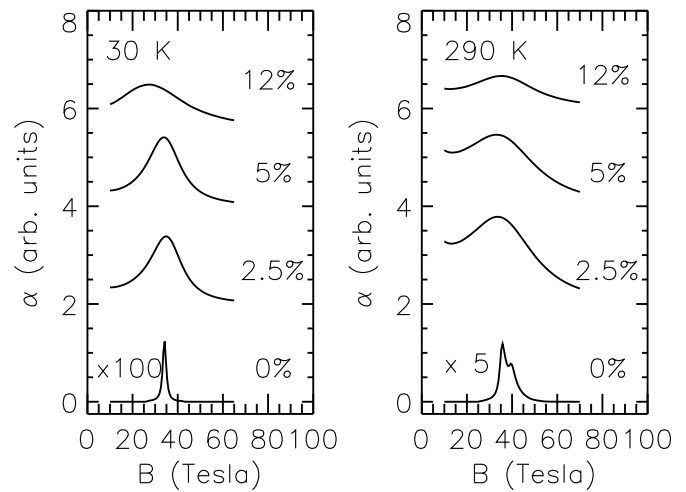}
\caption {Calculated CR absorption as a function of magnetic field
at 30 (a) and 290 K (b).
The curves were calculated based on the Pidgeon-Brown model
and the golden rule for absorption. The curves were broadened based
on the mobilities reported in Table I.}
\label{theorycr}
\end{figure}

Figures \ref{theorycr}(a) and \ref{theorycr}(b) show the calculated
CR absorption coefficient for electron-active
circularly polarized 10.6 $\mu$m light in the Faraday configuration as
a function of magnetic field at 30 K and 290 K, respectively. Densities
for each sample are given in Table I.  In the calculation, the curves
were broadened based on the mobilities of the samples.
The broadening used for $T$ = 30 K was 4 meV for 0\%, 40 meV
for 2.5\%, 40 meV for 5\% and 80 meV for 12\%.   For $T$ = 290 K, the
broadening used was 4 meV for 0\%, 80 meV for 2.5\%, 80 meV for 5\%
and 80 meV for 12\%.

At $T$ = 30 K, we see a shift in the CR peak as a
function of doping in agreement with Fig. \ref{figcr} (a).   For
$T$ = 290 K, we see the presence of two peaks in the pure InAs sample.
The second peak originates from the thermal population of the lowest
spin-down Landau level.  The peak does not
shift as much with doping as it did at low temperature.  This results
from the temperature dependence of the average Mn spin.
We believe that the Brillouin function used for calculating the average
Mn spin becomes inadequate at large $x$ and/or high
temperature due to its neglect of Mn-Mn interactions such as
pairing and clustering.

\section{Summary and conclusions}

We presented a theory for the electronic and magneto-optical
properties in $n$-type narrow gap In$_{1-x}$Mn$_{x}$As magnetic
semiconductor alloys in an ultrahigh external magnetic field, $B$,
oriented along [001]. 
We find several key results :  i) There is a shift in the cyclotron
resonance with Mn doping which is not predicted in simple models.   To
lowest order, this shift depends upon $(\alpha -\beta)$ and can be
thought of as  a renormalization of the energy gap in a magnetic field.
The value of the energy gap influences the amount of conduction-valence
band mixing.  ii) Even with no Mn doping, there is spin-splitting in
the cyclotron resonance which results from the nonparabolicity of the
conduction band in the narrow gap material.  iii) The relative heights
of the spin-split cyclotron absorption peaks can allow one to extract
information about carrier density.  iv) At high temperartures and high 
$(> 10\%)$ Mn concentrations, the calculated shift of the cyclotron
resonance 
is not as large as the experimentally observed shift. This probably
results from the inadequacy of the Brillouin
function used for calculating the average Mn spin at large $x$
and/or high temperature due to its neglect of Mn-Mn interactions such as
pairing and clustering 

In modeling the cyclotron resonance experiments
we used an 8 band Pidgeon-Brown model generalized to include the
wavevector dependence of the electronic states along $k_z$ as well
as the $s$-$d$ and $p$-$d$ exchange interactions with the
localized Mn $d$-electrons. Calculated conduction-band Landau
levels exhibit effective masses and $g$ factors that are strongly
dependent on temperature, magnetic field, Mn concentration $x$,
and $k_z$. At low temperatures and high $x$, the sign of the $g$
factor is positive and its magnitude exceeds 100. CR spectra are
computed using Fermi's golden rule and compared with
ultrahigh-magnetic-field ($>$ 50 T) CR experiments, which show
that the electron CR peak shifts sensitively with $x$. Detailed
comparison between theory and experiment allowed us to extract the
$s$-$d$ and $p$-$d$ exchange parameters, $\alpha$ and $\beta$. We
showed that not only $\alpha$ but also $\beta$ affects the
electron mass because of the strong interband coupling in this
narrow gap semiconductor. In addition, we derived analytical
expressions for effective masses and $g$ factors using the Kane
model, which indicates that ($\alpha - \beta$) is the crucial
parameter that determines the exchange interaction correction to
the cyclotron masses. These findings should be useful for
designing novel devices based on ferromagnetic semiconductors.

\appendix
\section{Landau Hamiltonian}
\label{A}

In this appendix, we write down the Landau contribution to the
Hamiltonian in the matrix eigenvalue problem (\ref{Schrodinger})
for an arbitrary Landau quantum number $n$.
The Landau Hamiltonian matrix can be obtained by taking the operator
form of the Landau Hamiltonian in Eq. (\ref{H_L}) and operating on the
envelope function wavefunction (\ref{Fn}) making use of the properties
of the creation and destruction operators $a^{\dagger}$ and $a$. The
resulting matrix is
\begin{equation}
H_L^{\prime} = \frac{\hbar^2}{m_0}\ \left[
\begin{array}{cc}
L_{a}^{\prime}        & L_{c}^{\prime} \\
L_{c}^{\dagger\prime} & L_{b}^{\prime}
\end{array}
\right]
\label{H_Lprime}
\end{equation}
where the submatrices are given by
\begin{widetext}
\begin{equation}
L_a^{\prime}=\left[
\begin{array}{cccc}
E^{\prime}+\frac{k^2}{2}+\gamma_4^{\prime\prime}(n-\frac{1}{2}) &
i V^{\prime} \sqrt{n-1} &
i V^{\prime} \sqrt{\frac{1}{3}n} &
  V^{\prime} \sqrt{\frac{2}{3}n}
\\
-i V^{\prime} \sqrt{n-1} &
(\gamma_2-\frac{\gamma_1}{2})k^2-\gamma_{12}(2n-3) &
-\gamma_{23} \sqrt{3n(n-1)} &
i \gamma_{23} \sqrt{6n(n-1)}
\\
-i V^{\prime} \sqrt{\frac{1}{3}n} &
-\gamma_{23} \sqrt{3n(n-1)} &
-(\gamma_2+\frac{\gamma_1}{2})k^2-\bar{\gamma}_{12}(2n+1) &
i\sqrt{2}(-\gamma_2k^2+\gamma_2^{\prime\prime}(n+\frac{1}{2}))
\\
 V^{\prime} \sqrt{\frac{2}{3}n} &
-i \gamma_{23} \sqrt{6n(n-1)} &
-i\sqrt{2}(-\gamma_2k^2+\gamma_2^{\prime\prime}(n+\frac{1}{2})) &
-\Delta^{\prime}-\gamma_1\frac{k^2}{2}
-\gamma_1^{\prime\prime}(n+\frac{1}{2})
\end{array}
\right]
\label{L_aprime}
\end{equation}
\begin{equation}
L_b^{\prime}=\left[
\begin{array}{cccc}
E^{\prime}+\frac{k^2}{2}+\gamma_4^{\prime\prime}(n+\frac{1}{2}) &
- V^{\prime} \sqrt{n+1} &
- V^{\prime} \sqrt{\frac{1}{3}n} &
 i V^{\prime} \sqrt{\frac{2}{3}n}
\\
-V^{\prime} \sqrt{n+1} &
(\gamma_2-\frac{\gamma_1}{2})k^2-\gamma_{12}(2n+3) &
-\gamma_{23} \sqrt{3n(n+1)} &
i \gamma_{23} \sqrt{6n(n+1)}
\\
- V^{\prime} \sqrt{\frac{1}{3}n} &
-\gamma_{23} \sqrt{3n(n+1)} &
-(\gamma_2+\frac{\gamma_1}{2})k^2-\bar{\gamma}_{12}(2n-1) &
i\sqrt{2}(-\gamma_2k^2+\gamma_2^{\prime\prime}(n-\frac{1}{2}))
\\
-i V^{\prime} \sqrt{\frac{2}{3}n} &
-i \gamma_{23} \sqrt{6n(n+1)} &
-i\sqrt{2}(-\gamma_2k^2+\gamma_2^{\prime\prime}(n-\frac{1}{2})) &
-\Delta^{\prime}-\gamma_1\frac{k^2}{2}
-\gamma_1^{\prime\prime}(n-\frac{1}{2})
\end{array}
\right]
\label{L_bprime}
\end{equation}
\begin{equation}
L_c^{\prime}=k\ \left[
\begin{array}{cccc}
0 &
0 &
V \sqrt{\frac{2}{3}}  &
i V \sqrt{\frac{1}{3}}
\\
0 &
0 &
i \gamma_3^{\prime} \sqrt{6(n-1)}  &
- \gamma_3^{\prime} \sqrt{n-1}
\\
-i V \sqrt{\frac{2}{3}} &
-i \gamma_3^{\prime} \sqrt{6(n+1)}  &
0 &
-3 \gamma_3^{\prime} \sqrt{n}
\\
-V \sqrt{\frac{1}{3}} &
\gamma_3^{\prime}  \sqrt{3(n+1)}  &
-3 \gamma_3^{\prime} \sqrt{n}  &
0
\end{array}
\right]
\label{L_cprime}
\end{equation}
\end{widetext}
where $k=k_z$ is the wavevector along the magnetic field direction,
and $n$ is the Landau quantum number for the manifold of states.
The submatrix $L_c^{\dagger\prime}$ is obtained by taking the
Hermitian adjoint of $L_c^{\prime}$.
In Eqs. (\ref{L_aprime}), (\ref{L_bprime}), and (\ref{L_cprime})
we make the following definitions:
\begin{equation}
E^{\prime}=\frac{m_0}{\hbar^2}E_g
\end{equation}
\begin{equation}
\Delta^{\prime}=\frac{m_0}{\hbar^2} \Delta
\end{equation}
\begin{equation}
V^{\prime}= \frac{m_0}{\hbar^2}\frac{V}{\lambda}
\end{equation}
\begin{equation}
\gamma_i^{\prime}=\frac{\gamma_i}{\lambda}\ \ (i=1 \dots 4)
\end{equation}
\begin{equation}
\gamma_i^{\prime\prime}=\frac{\gamma_i}{\lambda^2}\ \ (i=1 \dots 4)
\end{equation}
\begin{equation}
\gamma_{ij}=\frac{1}{\lambda^2}\frac{\gamma_i+\gamma_j}{2}
\ \ (i,j=1 \dots 4)
\end{equation}
\begin{equation}
\bar{\gamma}_{ij}=\frac{1}{\lambda^2}\frac{\gamma_i-\gamma_j}{2}
\ \ (i,j=1 \dots 4)
\end{equation}
Here $E_g$ and $\Delta$ are the band gap and spin-orbit splitting
energies, $\lambda$ is the magnetic length defined in
Eq. (\ref{lambda}), $V$ is the Kane momentum matrix element defined
in Eq. (\ref{V}), and the $\gamma_i's$ are the effective mass
parameters defined in Eq. (\ref{APQLM}).
The total Hamiltonian, $H_n$, to be diagonalized in the eigenvalue
equation (\ref{Schrodinger}) is the sum of the Landau Hamiltonian
matrix (\ref{H_Lprime}) and the Zeeman and exchange Hamiltonians
(\ref{H_Z}) and (\ref{H_Mn}).

We note from Eq. (\ref{L_cprime}) that the submatrix $L_c^{\prime}$
is proportional to $k$ and so vanishes at $k=0$. In this limit,
$H_n$, is block diagonal with respect to the upper and lower Bloch
basis sets defined in Eqs. (\ref{upperset}) and (\ref{lowerset}).

\section{Optical matrix elements}
\label{B}

In this appendix we write down the momentum matrix elements used in the
computation of optical matrix elements in Eq. (\ref{EdotP}). For the
Bloch basis states defined in Eqs (\ref{upperset}) and (\ref{lowerset}),
the matrix elements for the momentum operators $p_x$, $p_y$, and
$p_z$ are given by

\begin{widetext}
\begin{equation}
P_x=\left[
\begin{array}{cccccccc}
0 &
iV\sqrt{\frac{1}{2}} &
iV\sqrt{\frac{1}{6}} &
V\sqrt{\frac{1}{3}} & 0 & 0 & 0 & 0 \\
-iV\sqrt{\frac{1}{2}} & 0 & 0 & 0 & 0 & 0 & 0 & 0 \\
-iV\sqrt{\frac{1}{6}} & 0 & 0 & 0 & 0 & 0 & 0 & 0 \\
  V\sqrt{\frac{1}{3}} & 0 & 0 & 0 & 0 & 0 & 0 & 0 \\
0 & 0 & 0 & 0 & 0 &
-V\sqrt{\frac{1}{2}} &
-V\sqrt{\frac{1}{6}} &
iV\sqrt{\frac{1}{3}} \\
0 & 0 & 0 & 0 &  -V\sqrt{\frac{1}{2}} & 0 & 0 & 0 \\
0 & 0 & 0 & 0 &  -V\sqrt{\frac{1}{6}} & 0 & 0 & 0 \\
0 & 0 & 0 & 0 & -iV\sqrt{\frac{1}{3}} & 0 & 0 & 0
\end{array}
\right]
\end{equation}
\begin{equation}
P_y=\left[
\begin{array}{cccccccc}
0 &
 -V\sqrt{\frac{1}{2}} &
  V\sqrt{\frac{1}{6}} &
-iV\sqrt{\frac{1}{3}} & 0 & 0 & 0 & 0 \\
 -V\sqrt{\frac{1}{2}} & 0 & 0 & 0 & 0 & 0 & 0 & 0 \\
  V\sqrt{\frac{1}{6}} & 0 & 0 & 0 & 0 & 0 & 0 & 0 \\
 iV\sqrt{\frac{1}{3}} & 0 & 0 & 0 & 0 & 0 & 0 & 0 \\
0 & 0 & 0 & 0 & 0 &
iV\sqrt{\frac{1}{2}} &
-iV\sqrt{\frac{1}{6}} &
 -V\sqrt{\frac{1}{3}} \\
0 & 0 & 0 & 0 & -iV\sqrt{\frac{1}{2}} & 0 & 0 & 0 \\
0 & 0 & 0 & 0 &  iV\sqrt{\frac{1}{6}} & 0 & 0 & 0 \\
0 & 0 & 0 & 0 &  -V\sqrt{\frac{1}{3}} & 0 & 0 & 0
\end{array}
\right]
\end{equation}
\begin{equation}
P_z=\left[
\begin{array}{cccccccc}
0 & 0 & 0 & 0 & 0 & 0 & V\sqrt{\frac{2}{3}} & iV\sqrt{\frac{1}{3}} \\
0 & 0 & 0 & 0 & 0 & 0 & 0 & 0 \\
0 & 0 & 0 & 0 & -iV\sqrt{\frac{2}{3}} & 0 & 0 & 0 \\
0 & 0 & 0 & 0 &  -V\sqrt{\frac{1}{3}} & 0 & 0 & 0 \\
0 & 0 & iV\sqrt{\frac{2}{3}} & -V\sqrt{\frac{1}{3}} & 0 & 0 & 0 & 0 \\
0 & 0 & 0 & 0 & 0 & 0 & 0 & 0 \\
  V\sqrt{\frac{2}{3}} & 0 & 0 & 0 & 0 & 0 & 0 & 0 \\
-iV\sqrt{\frac{1}{3}} & 0 & 0 & 0 & 0 & 0 & 0 & 0
\end{array}
\right]
\end{equation}
\end{widetext}
where $V$ is the Kane matrix element defined in Eq. (\ref{V}).

\begin{acknowledgments}
This work was supported by DARPA through grant No. MDA972-00-1-0034, the
National Science Foundation through grant DMR 9817828,  and
the NEDO International Joint Research Program.
\end{acknowledgments}

\bibliography{paper}

\end{document}